\documentclass[a4paper,fleqn,usenatbib]{mnras}

\usepackage[T1]{fontenc}
\usepackage{ae,aecompl}

\usepackage[pdftex]{graphicx}
\usepackage{hyperref}
\usepackage{natbib}
\usepackage{amssymb}
\usepackage{amsmath}
\usepackage{amstext}
\usepackage{lscape}
\usepackage{xcolor}
\usepackage[varg]{txfonts}	
\usepackage{microtype}		
\usepackage{multirow}
\usepackage{rotating}
\usepackage{ulem}
\usepackage[autolanguage]{numprint}
\npfourdigitnosep

\newcommand\hbeta{\ensuremath{\mathrm{H}\beta}}
\newcommand\halpha{\ensuremath{\mathrm{H}\alpha}}
\newcommand\elec{\ensuremath{_{\mathrm{e}}}}

\newcommand{\te}{\ensuremath{T\elec}}
\newcommand{\nel}{\ensuremath{n\elec}}

\newcommand{\chb}{\ensuremath{c(\hbeta)}}

\newcommand\ioni[2]{\ensuremath{\mathrm{#1}^{#2}}}           

\newcommand\unit[1]{\ensuremath{\mathrm{#1}}}

\newcommand\cmc{\unit{cm^{-3}}}

\title[The Fe/Ni ratio in ionized nebulae]{The Fe/Ni ratio in ionized nebulae: clues on dust depletion patterns}

\author[G. Delgado-Inglada et al.]{
G. Delgado-Inglada,$^{1}$\thanks{E-mail: gdelgado@astro.unam.mx (GDI)}
A. Mesa-Delgado,$^{2}$
J. Garc\'ia-Rojas,$^{3,4}$
M. Rodr\'iguez,$^{5}$
\newauthor{and C. Esteban,$^{3,4}$}
\\
$^{1}$Instituto de Astronom\'ia, Universidad Nacional Aut\'onoma de M\'exico, Apdo. Postal 70264, M\'ex. D. F., 04510, Mexico\\
$^{2}$Instituto de Astrof\'isica, Facultad de F\'isica, Pontificia Universidad Cat\'olica de Chile, Av.~Vicu\~na Mackenna 4860, 782-0436 Macul, Santiago, Chile\\
$^{3}$Instituto de Astrof\'isica de Canarias, E-38200 La Laguna, Tenerife, Spain\\
$^4$Universidad de La Laguna. Departamento de Astrof\'isica, E-38206 La Laguna, Tenerife, Spain\\
$^5$Instituto Nacional de Astrof\'isica, \'Optica y Electr\'onica (INAOE), Apdo Postal 51 y 216, 72000, Puebla, Mexico\\}

\date{Accepted 2015 December 17. Received 2015 December 03; in original form 2015 October 13}
\pubyear{2015}

\begin{document}
\label{firstpage}
\pagerange{\pageref{firstpage}--\pageref{lastpage}}
\maketitle

\begin{abstract}
We perform a homogeneous analysis of the Fe/Ni abundance ratio in eight Galactic planetary nebulae (PNe) and three Galactic \ion{H}{ii} regions that include the Orion nebula, where we study four nebular zones and one shocked region. We use [\ion{Fe}{ii}], [\ion{Fe}{iii}], and [\ion{Ni}{iii}] lines, and ionization correction factors (ICFs) that account for the unobserved ions. We derive an ICF for nickel from an extensive grid of photoionization models. We compare our results with those derived by other authors for 16 neutral clouds in the solar neighbourhood with available Fe/Ni ratios in the literature. 

We find an excellent agreement between the ionized nebulae and the diffuse clouds, with both types of regions showing a clear correlation between the Fe/Ni ratios and the iron and nickel depletion factors. The trend shows that the objects with a relatively low depletion have near solar Fe/Ni ratios whereas at higher depletions the Fe/Ni ratio increases with the depletion. Our results confirm that, compared to iron atoms, nickel ones are more efficiently stuck to the dust grains in ambients where dust formation or growth have been more efficient. 
\end{abstract}


\section{Introduction} \label{intro}

Dust plays a major role in many galactic aspects, from the formation and evolution of stars to the chemical, thermodynamic, and dynamic evolution of the interstellar medium \citep[ISM; see, e.g.,][]{whi03}. Our present knowledge establishes that dust grains are mainly formed in the latest stages of stellar evolution, during the giant and supergiant phases and in nova or supernova events. In particular, the cool and dense atmospheres of asymptotic giant branch (AGB) stars have been identified as the main sources of stardust in our Galaxy \citep{dra09}. After being formed, dust grains are injected into the ISM, where they are subject to many processes that may destroy them, returning the elements to the gas \citep{whi03, jones11}. In the densest regions of the ISM dust grains grow by accretion and coagulation which are expected to be the main processes that determine the amount of interstellar dust \citep{dra90, dra09, hir12}.   

The presence of dust grains in ionized nebulae is revealed by direct images and by their rich infrared spectra. An indirect evidence is provided by the observed element depletions. The depletion factor of a given element, X, refers to the observed under-abundance with respect to a given reference, and is defined as:
\begin{equation}
[{\rm X}/{\rm H}] = \log({\rm X}/{\rm H}) - \log({\rm X}/{\rm H})_\odot.
\end{equation}
The solar abundances are generally used as reference of cosmic abundances in the solar neighbourhood because the Sun is the object for which we can more accurately calculate the abundances of a greater number of elements. If all the atoms of the element X are in the gas phase, the measured abundance is expected to be similar to the reference (i.e.~solar) value. The observed deficiency is interpreted as due to the incorporation of the atoms to the solid phase. The study of depletions from the emission line spectra of planetary nebulae (PNe) and \ion{H}{ii} regions provides clues about the life cycle of dust grains. Whereas in PNe the dust is fresh, recently formed and ejected in the previous AGB phase, in \ion{H}{ii} regions the dust is processed dust, formed by grains that were originally inside the associated molecular cloud where they could grow. Hence, PNe provide information about the efficiency of dust formation in AGB stars and \ion{H}{ii} regions allow us to investigate the efficiency of dust growth in the ISM. Besides, in both cases there might be some destruction or processing of the dust. 

Refractory elements, such as Al, Ca, Fe, Ni, and Si are mostly condensed into dust grains and some of them seem to strongly resist their re-incorporation to the gas phase \citep{bar78, jen09}. Therefore, these elements can be used to investigate how their depletions vary from one object to another and among different environments, providing clues on dust formation and evolution mechanisms in different phases of the ISM. However, in ionized nebulae there are several problems involved in the abundance calculation of refractory elements. First, the number of deep and high resolution spectra where the emission lines of these elements can be reliable measured is scarce. Second, the atomic data needed for the calculations are not always available. And third, there are no ionization correction factors (ICFs) in the literature to derive the total abundances of some of these elements. \cite{delgadoingladaetal09, del14a} performed a detailed analysis of the iron abundances in a large sample of PNe and \ion{H}{ii} regions from our Galaxy. They found that all the objects in their sample have more than 90 per cent of their iron atoms in the dust phase, but there is a high dispersion in the results, especially for the PNe. They also found that PNe with carbon-rich grains (those with infrared features associated with SiC or MgS) tend to have higher iron depletions than PNe with oxygen-rich dust (i.e.~silicates). 

Nickel has physical and chemical properties similar to those of iron (for example, they have a similar condensation temperature, $T_c$ $\sim$ 1300 K). In a very simplistic scenario we expect both elements to form the same type of dust compounds and to be incorporated into the dust grains in a similar proportion \citep{fie74, hen84}. In fact they are found together in planet cores and meteorites in the solar system \citep[see, e.g.,][]{lod10, mor13}. In the neutral ISM, \citet{jen09} found a similar depletion pattern for both elements. In stars, nickel abundances follow iron abundances for a wide range in metallicity \citep[see, e.g.,][]{red03, red06, ben14}. This implies that any difference between the Fe/Ni abundance ratio in stars and in the ISM can be attributed to dust production and destruction processes. 

The first attempt to determine the gaseous abundance of Ni in an ionized gas was performed in the Crab and Orion nebulae by \cite{hen84, ost92}. They used estimates of the atomic data of Ni ions, and considered both [\ion{Ni}{ii}] and [\ion{Ni}{iii}] lines to calculate the ionic abundances. Since then, the number of nickel abundance determinations in ionized nebulae available in the literature is very scarce. Some recent calculations of the nickel abundances in ionized nebulae have been performed, for example, by \citet{zhangliu06, mesadelgadoetal09b}; and \citet{gar13}. These authors used the [\ion{Ni}{ii}] and [\ion{Ni}{iii}] lines, and ICFs based on the similarity between ionization potentials of Fe$^+$ (16.2 eV) and Ni$^+$ (18.2 eV) in the first paper, and of Ni$^{3+}$ (35.2 eV) and O$^{++}$ (35.1 eV) in the other two papers to calculate the nickel abundances. The derived values are uncertain because the [\ion{Ni}{ii}] lines are likely affected by fluorescence \citep{lucy95} and because they were computed using ICF schemes that are not necessarily correct (see our discussion in Section~\ref{sec:totab}). 

The present paper aims to investigate both the iron and nickel abundances in the ionized phase of the ISM. To achieve this goal, we looked in the literature for the deepest spectroscopic observations of ionized nebulae, PNe and \ion{H}{ii} regions, containing reliable detections of [\ion{Fe}{iii}] and [\ion{Ni}{iii}] emission lines (Section~\ref{sample}). We decided to avoid, as far as we can, the use of Fe$^+$ and Ni$^+$ to determine the total abundances because these two ions are minor contributors to the total abundances of iron and nickel, and because the [\ion{Ni}{ii}] lines and some of the [\ion{Fe}{ii}] lines are affected by fluorescence. A homogeneous analysis was performed in all the objects using the same collection of updated atomic data, re-computing physical conditions (Section~\ref{sec:phycond}), ionic abundances (Section~\ref{sec:ionab}) and total abundances (Section~\ref{sec:totab}). Nickel abundances were calculated from the [\ion{Ni}{iii}] lines and the new ICF scheme obtained here from an extensive grid of photoionization models. Iron abundances were derived from the [\ion{Fe}{ii}] lines and the [\ion{Fe}{iii}] lines, in combination with the ICFs derived by \citet{rodriguezrubin05}. This approach constitutes the best option for determining the total gaseous abundance of nickel and iron in ionized nebulae. The results are discussed in Section~\ref{discussion} and the main conclusions of the work are summarized in Section~\ref{conclu}.
  
\section{The Sample}\label{sample}
The sample studied here consists of all the ionized nebulae in the literature with detections of [\ion{Ni}{iii}] and  [\ion{Fe}{iii}] lines. Within the sample there are eight PNe: Mz~3 \citep{zhangliu02}, IC~418 \citep{sharpeeetal03}, NGC~7027 \citep{zhangetal05}, NGC~7009 \citep{fangliu11}, Cn~1-5, He~2-86, M~1-30, and M~1-32 \citep{garciarojasetal12}. In Mz~3, we only considered the data for the extended inner lobes (labeled as ``P2'' in the original paper) since the central core region shows a complex density structure, with evidence of a density gradient \citep{zhangliu02}, which hampers the calculation of reliable physical conditions and abundances.

The sample also contains three \ion{H}{ii} regions: NGC~3576 \citep{garciarojasetal04}, the Orion nebula \citep{estebanetal04, mesadelgadoetal09b}, and M~8 \citep{garciarojasetal07}. From observations under the ESO programs ID 068.C-0149(A) and 070.C-0008(A), our analysis makes use of unpublished data of two slit positions in the Orion nebula, named here as ``P1'' and ``Bar''. These positions were observed to complement the deep spectroscopic study of the Orion nebula presented by \cite{estebanetal04}, labelled here as ``P2''.  
The ``P1''  position covers approximately the same area as position 1 observed by \citet{est98}, located close to the centre of the nebula, at 45 arcsec N of the star $\theta^1$~Ori~C. The ``Bar'' position is centered in one of the brightest zones of the Orion Bar, at 24 arcsec N and 12 arcsec W of $\theta^2$~Ori~A. In both cases, the slit was oriented east-west and had a size of 10 $\times$ 3 arcsec$^2$. For  ``P1'' and ``Bar'' the reduction process, including the extractions of the one-dimensional spectra, as well as the flux measurements and reddening correction were performed following the same strategy described by \cite{estebanetal04}. In Table~\ref{orionlines}, we present the de-reddened intensity ratios of the emission lines relevant to our analysis for these new observations. We also consider here the high-quality spectroscopic data retrieved from observations of the Herbig-Haro 202 (HH~202) object in Orion \citep{mesadelgadoetal09b}. The two kinematic components observed in HH~202 were considered in the sample: one associated with the emission from the nebular background of Orion --HH~202~N-- and another one emitting from the irradiated, high-velocity gas flow --HH~202~S.
\begin{table}
\centering
\begin{minipage}{80mm}
\caption{Emission line list of Orion observations, named as P1 and Bar. De-reddened intensity ratios of the lines relevant to this analysis are in units of I(\hbeta) = 100.}
\label{orionlines}
\begin{tabular}{ccccc}
\hline  
$\lambda_{air}$ (\AA) & Ion & Mult. & Orion (P1) & Orion (Bar)  \\     
\hline
3322.54 & [\ion{Fe}{iii}] & 5F & $0.06\pm0.02$ & $0.17\pm0.03$ \\
3726.03 & [\ion{O}{ii}]   & 1F & $58\pm2$ & $164\pm6$ \\ 	
3728.82 & [\ion{O}{ii}]   & 1F & $30\pm1$ & $87\pm3$ \\
3970.07 & \ion{H}{i}	  & H7 & $16\pm1$ & $16\pm1$ \\
4008.36 & [\ion{Fe}{iii}] & 4F & $0.025\pm0.005$ & $0.055\pm0.003$ \\
4101.74 & \ion{H}{i}	  & H$\delta$ & $25\pm1$ & $25\pm1$ \\
4340.47 & \ion{H}{i}	  & H$\gamma$ & $44\pm1$ & $46\pm1$ \\
4363.21 & [\ion{O}{iii}]  & 2F & $1.18\pm0.04$ & $0.63\pm0.02$ \\	
4471.48 & \ion{He}{i}     & 14 & $4.5\pm0.1$ & $4.2\pm0.1$ \\
4596.84 & [\ion{Ni}{iii}] & $^3$F-$^1$G & -- & $0.009\pm0.002$ \\
4607.13 & [\ion{Fe}{iii}] & 3F & $0.050\pm0.005$ & $0.078\pm0.007$ \\
4658.10 & [\ion{Fe}{iii}] & 3F & $0.64\pm0.02$ & $1.30\pm0.04$ \\ 
4667.01 & [\ion{Fe}{iii}] & 3F & $0.035\pm0.003$ & $0.051\pm0.003$ \\
4701.53 & [\ion{Fe}{iii}] & 3F & $0.20\pm0.01$ & $0.37\pm0.01$ \\
4711.37 & [\ion{Ar}{iv}]  & 1F & $0.035\pm0.004$ & $0.007\pm0.002$\\
4733.93 & [\ion{Fe}{iii}] & 3F & $0.075\pm0.006$ & $0.14\pm0.01$ \\
4740.16 & [\ion{Ar}{iv}]  & 1F & $0.045\pm0.005$ & $0.007\pm0.001$ \\
4754.83 & [\ion{Fe}{iii}] & 3F & $0.12\pm0.01$ & $0.24\pm0.01$ \\
4769.60 & [\ion{Fe}{iii}] & 3F & $0.07\pm0.01$ & $0.13\pm0.01$ \\
4777.68 & [\ion{Fe}{iii}] & 3F & $0.039\pm0.004$ & $0.066\pm0.004$ \\
4861.03 & \ion{H}{i}	  & \hbeta & $100\pm3$ & $100\pm3$ \\
4881.00 & [\ion{Fe}{iii}] & 2F & $0.29\pm0.01$ & $0.57\pm0.02$\\
4924.50 & [\ion{Fe}{iii}] & 2F & $0.028\pm0.004$ & $0.025\pm0.002$ \\
4930.50 & [\ion{Fe}{iii}] & 1F & $0.023\pm0.004$ & $0.039\pm0.003$ \\ 
4958.91 & [\ion{O}{iii}]  & 1F & $120\pm4$ & $58\pm2$ \\
4985.90 & [\ion{Fe}{iii}] & 2F & $0.015\pm0.003$ & $0.042\pm0.007$ \\
4987.20 & [\ion{Fe}{iii}] & 2F & $0.052\pm0.005$ & $0.11\pm0.01$ \\
5006.84 & [\ion{O}{iii}]  & 1F & $358\pm11$ & $174\pm5$ \\
5011.30 & [\ion{Fe}{iii}] & 1F & $0.08\pm0.01$ & $0.15\pm0.01$ \\
5270.40 & [\ion{Fe}{iii}] & 1F & $0.34\pm0.01$ & $0.66\pm0.02$ \\
5412.00 & [\ion{Fe}{iii}] & 1F & $0.031\pm0.004$ & $0.059\pm0.003$ \\
5517.71 & [\ion{Cl}{iii}] & 1F & $0.36\pm0.02$ & $0.46\pm0.02$ \\
5537.88 & [\ion{Cl}{iii}] & 1F & $0.52\pm0.02$ & $0.58\pm0.02$ \\ 
5754.64 & [\ion{N}{ii}]	  & 3F & $0.49\pm0.02$ & $1.32\pm0.04$ \\
5875.64 & \ion{He}{i}	  & 11 & $14\pm1$ & $13\pm1$ \\
6000.20 & [\ion{Ni}{iii}] & 2F & $0.012\pm0.004$ & $0.015\pm0.002$ \\
6401.50 & [\ion{Ni}{iii}] & 2F & $0.007\pm0.002$ & $0.008\pm0.002$ \\
6533.80 & [\ion{Ni}{iii}] & 2F & -- & $0.022\pm0.003$ \\
6548.03 & [\ion{N}{ii}]	  & 1F & $11\pm1$ & $33\pm1$ \\ 
6562.82 & \ion{H}{i}	  & \halpha & $289\pm10$ & $283\pm10$ \\
6583.41 & [\ion{N}{ii}]	  & 1F & $33\pm1$ & $103\pm4$ \\
6678.15 & \ion{He}{i}	  & 46 & $3.3\pm0.1$ & $3.4\pm0.1$ \\
6682.20 & [\ion{Ni}{iii}] & 2F & $0.004\pm0.001$ & $0.008\pm0.001$ \\	
6716.47 & [\ion{S}{ii}]	  & 2F & $1.9\pm0.1$ & $7.0\pm0.3$ \\
6730.85 & [\ion{S}{ii}]	  & 2F & $3.1\pm0.1$ & $12\pm1$ \\
6797.10 & [\ion{Ni}{iii}] & 2F & -- & $0.0030\pm0.0005$ \\	
7155.16 & [\ion{Fe}{ii}] & 14F & $0.054\pm0.004$ & $0.170\pm0.003$ \\
7889.90 & [\ion{Ni}{iii}] & 1F & $0.040\pm0.003$ & $0.056\pm0.003$ \\	
9229.01 & \ion{H}{i}	  & P9 & $2.1\pm0.2$ & $2.2\pm0.1$ \\
9545.97 & \ion{H}{i}	  & P8 & $2.7\pm0.2$ & $2.9\pm0.2$ \\

\multicolumn{3}{c}{F(\hbeta) ($10^{-11}$~erg~cm$^{-2}$~s$^{-1}$)}  & $1.59\pm0.05$ & $1.02\pm0.03$ \\
\multicolumn{3}{c}{\chb}& $0.64\pm0.04$ & $0.44\pm0.04$ \\
\hline
\end{tabular}
\end{minipage}
\end{table}
\section{The analysis}\label{analysis}
With the aim of performing a homogeneous analysis of the nebulae, we recomputed their physical conditions and chemical abundances making use of what, according to our experience \citep[see, e.g.,][]{del14a, est15}, is the best set of atomic dataset available at the moment (see Table~\ref{atomic}). All calculations were carried out with the software {\sc PyNeb} \citep{luridianaetal15}, a Python package for the computation of physical conditions and chemical abundances in ionized nebulae.

\begin{table}
   \centering
   \begin{minipage}{90mm}
    \caption{Atomic dataset.}
    \label{atomic}
    \begin{tabular}{lll}
    \hline  
         & Transition  & Collision\\
         Ion & Probabilities  &Strengths\\
    \hline
N$^+$ & \cite{froesefischertachiev04} & \cite{tayal11} \\
O$^+$ & \cite{froesefischertachiev04}  & \cite{kisieliusetal09}\\
O$^{++}$&  \cite{froesefischertachiev04} & \cite{storeyetal14}\\
S$^+$&  \cite{mendozazeippen82a} & \cite{TayalZatsarinny10} \\
Cl$^{++}$& \cite{mendozazeippen82a}& \cite{ButlerZeippen89} \\
Ar$^{3+}$& \cite{mendozazeippen82a}& \cite{Ramsbottometal97}  \\
Fe$^{++}$&  \cite{quinet96} & \cite{zhang96} \\
                     & \cite{Johanssonetal00} & \\
Ni$^{++}$ & \cite{bautista01}& \cite{bautista01} \\
         \hline
    \end{tabular}
   \end{minipage}
 \end{table}

We started from the de-reddened intensity ratios with respect to \hbeta, as well as their uncertainties, provided by the different authors. The uncertainties in the intensity ratios are not provided for NGC~7027 and Mz~3. For these objects we adopted typical uncertainties in the lines involved: 5 per cent for intensity ratios $I(\lambda)/I($\hbeta$)>0.2$ (in units of $I($\hbeta$)=100$); 10 per cent for those ratios with $0.1<I(\lambda)/I($\hbeta$)<0.2$; 20 per cent for $0.05<I(\lambda)/I($\hbeta$)<0.1$; and 30 per cent for lines with $I(\lambda)/I($\hbeta$)<0.05$ \citep[see e.g.][]{tsamisetal03b}.

The physical conditions and the ionic abundances were obtained by propagating the uncertainties in the intensity ratios through the implementation of a Monte Carlo procedure. We generated 200 random values for each intensity ratio using a Gaussian distribution centered in the observed value and with a sigma equal to the flux uncertainty. Then, we computed all the quantities (physical conditions, ionic, and total abundances) for each Montecarlo run. We adopted as the final values the ones derived from the observed intensity ratios (except for the density, see \ref{sec:phycond}), and as the final uncertainties associated with each quantity the ones derived from the 16 and 84 percentiles, that define a confidence interval of 68  per cent. We checked that for a higher number of Monte Carlo simulations the uncertainties in the computed quantities remain practically the same.
\subsection{Physical conditions}\label{sec:phycond}
Electron densities (\nel) can be calculated from several diagnostic ratios based on collisionally excited lines. Some of the most commonly used are [\ion{O}{ii}]~$\lambda$3726/$\lambda$3729, [\ion{S}{ii}]~$\lambda$6716/$\lambda$6731, [\ion{Cl}{iii}]~$\lambda$5517/$\lambda$5531, and [\ion{Ar}{iv}]~$\lambda$4711/$\lambda$4740. Table~\ref{phycond} displays the values obtained with each of the aforementioned diagnostic ratios. Looking at this table we can see that most of the nebulae have densities $\gtrsim5000$ \cmc, and exceed 20\,000 \cmc\ in a few cases: He~2-86, NGC~7027, and Orion (HH~202 S). 

We decided to calculate for each Monte Carlo run the average of the densities derived with the available diagnostic ratios among the four already mentioned. Then, we adopted as the final density the median value of the density distribution obtained for each object and as the final uncertainties the ones derived from the 16 and 84 percentiles. These densities were then used to compute the electron temperatures (\te) from the line ratios [\ion{N}{ii}]~$\lambda$5755/($\lambda$6548+$\lambda$6583) and [\ion{O}{iii}]~$\lambda$4363/($\lambda$4959+$\lambda$5007). The final adopted physical conditions for each object are shown in columns 6, 7, and 8 of Table~\ref{phycond}. In section~\ref{discussion} we discuss how the adoption of a different approach to define the representative physical conditions of the nebulae does not change in a significant way our conclusions.

\begin{table*}
\centering
\begin{minipage}{145mm}
\caption{Physical conditions$^{a}$.}\label{phycond}
\begin{tabular}{lr@{}lr@{}lr@{}lr@{}lr@{}lr@{}lr@{}l}
\hline  
& \multicolumn{10}{c}{\nel~(\cmc)} & \multicolumn{4}{c}{\te~(K)} \\ 
& \multicolumn{2}{c}{[\ion{O}{ii}]}  &  \multicolumn{2}{c}{[\ion{S}{ii}]} &  \multicolumn{2}{c}{[\ion{Cl}{iii}]} &  \multicolumn{2}{c}{[\ion{Ar}{iv}]} &   \multicolumn{2}{c}{Adopted} &  \multicolumn{2}{c}{[\ion{N}{ii}]} &  \multicolumn{2}{c}{[\ion{O}{iii}]}  \\     
\hline
\multicolumn{14}{l}{\bf Planetary nebulae}\\
Cn~1-5          & $5300$ & $^{+2400}_{-1900}$        & $4900$ & $^{+2500}_{-1700}$     & $3400$ & $^{+700}_{-600}$          & $10000$ & $^{+5100}_{-4600}$   & $5900$ & $^{+2000}_{-1200}$             & $8600$ & $^{+200}_{-300}$   & $8800$ & $\pm200$ \\
He~2-86        & $15200$ & $^{+15300}_{-7900}$    & $19900$ & $^{+4800}_{-13300}$ & $17400$ & $^{+3900}_{-3200}$    & $35400$ & $^{+7300}_{-4600}$   & $22000$ & $^{+5900}_{-3300}$ & $10400$ & $\pm600$  & $8400$ & $^{+100}_{-200}$ \\
IC~418         & $15800$ & $^{+14000}_{-7200}$     & $19000$ & $^{+8900}_{-10000}$  & $10200$ & $\pm2500$                  & $4900$ & $^{+14300}_{-2900}$    & $12500$ & $^{+8700}_{-3200}$                & $9300$ & $\pm400$   & $8800$ & $^{+100}_{-200}$ \\
M~1-30         & $4800$ & $^{+2900}_{-1300}$        & $6800$ & $^{+7500}_{-3600}$      & $6200$ & $^{+1200}_{-1100}$        & $\ldots$ &                         & $5900$ & $^{+3600}_{-10100}$  & $7000$ & $^{+200}_{-300}$               & $6600$ & $\pm200$ \\
M~1-32         & $9700$ & $^{+10100}_{-4600}$      & $10900$ & $^{+7900}_{-7200}$    & $11700$ & $^{+3000}_{-2800}$     & $\ldots$ &                        & $10800$ & $^{+7500}_{-3200}$ & $9000$ & $\pm600$               & $9400$ & $\pm200$ \\
Mz~3 (P2)    & $3900$ & $^{+1700}_{-800}$         & $5600$ & $^{+2500}_{-1700}$      & $9400$ & $^{+3000}_{-2000}$     & $\ldots$  &                        & $6300$ & $^{+1900}_{-800}$     & $6800$ & $\pm100$              & $13300$ & $\pm300$ \\
NGC~7009   & $5900$ & $\pm100$                        & $4200$ & $\pm200$                      & $3400$ & $\pm200$                      & $4200$ & $\pm300$         & $4400$ & $\pm100$                  & $12000$ & $\pm300$             & $9900$ &  $\pm200$ \\
NGC~7027   & $\ldots$  &                                      & 52300&:                    & $49500$ & $^{+31500}_{-14400}$ & $51200$ & $^{+7600}_{-5900}$   & $51000$ & $^{+7300}_{-13000}$  & $12800$ & $^{+1500}_{-800}$    & $12500$ & $^{+400}_{-300}$ \\
\multicolumn{14}{l}{\bf \ion{H}{ii} regions}\\
M~8             & $1600$ & $^{+600}_{-500}$           & 1600 & $\pm200$    & 1600 & $^{+300}_{-200}$                  & 2000 & $^{+6500}_{-900}$     & 1700 & $^{+1200}_{-200}$                 & 8300 & $^{+100}_{-200}$             & 8100 & $\pm100$ \\
NGC~3576  & $1700$ & $\pm300$           & 1400 & $\pm400$     & 2500 & $^{+600}_{-500}$      & 2900 & $^{+1700}_{-1200}$   & 2100 & $^{+500}_{-300}$     & 8600 & $\pm200$             & 8500 & $\pm100$\\
Orion (Bar)  & $4200$ & $^{+700}_{-600}$           & 4200 & $^{+800}_{-600}$     & 3600 & $\pm400$      & 3100 & $^{+4200}_{-1900}$   & 3800 & $^{+1000}_{-500}$    & 9100 & $\pm200$              & 8400 & $\pm100$\\
Orion (HH~202 N) & $2900$ & $^{+900}_{-600}$  & 2400 & $^{+900}_{-500}$     & 1800 & $^{+900}_{-700}$      & 4100 & $^{+11600}_{-2700}$ & 2800 & $^{+2600}_{-700}$     & 9500 & $^{+300}_{-500}$             & 8100 & $\pm200$ \\
Orion (HH~202 S) & $29700$ & $^{+20200}_{-16200}$ & 31600   & $^{+8500}_{-21900}$          & 11400 & $^{+3300}_{-2700}$ & $\ldots$  &           & 24200 & $^{+5400}_{-12500}$ & 9200 & $^{+1100}_{-400}$ & 8700 & $\pm200$ \\
Orion (P1)   & $4800$ & $^{+1400}_{-1000}$         & 3700 & $^{+1000}_{-700}$  & 5100 & $\pm700$  & 5500 & $^{+2100}_{-1800}$    & 4800 & $^{+900}_{-600}$    & 9500 & $\pm200$              & 8200 & $\pm100$ \\
Orion (P2)   & $6800$ & $^{+2900}_{-1600}$       & 6000 & $^{+3000}_{-1900}$ & 5900 & $\pm500$      & 4900 & $\pm900$     & 5900 & $^{+1400}_{-700}$                & 10200 & $^{+200}_{-300}$ & 8300 & $\pm100$\\
\hline
\end{tabular}
\begin{description}
\item $^a$ Colons ``:'' indicate unreliable values.
\end{description}   
\end{minipage}
\end{table*}

In general, the physical conditions derived here are in agreement with the ones computed in the original works. Small differences are found in the density determinations and \te([\ion{O}{iii}]), mainly arising from the use of different atomic data. The largest differences are found in \te([\ion{N}{ii}]) for He~2-86, M~1-30, and NGC~7009 (up to 2500 K). The reason is that we have not taken into account the contribution of recombination to the [\ion{N}{ii}] $\lambda$5755 line. This correction can be estimated from the expression derived by \cite{liu00}, but it is somewhat uncertain since it is based on the N$^{++}$ abundance that can be estimated either from collisionally excited lines or recombination lines, leading to different results, and thus, we prefer not to consider it. We estimated that the effect of this correction in the final Fe/Ni ratio of NGC~7009 (the most extreme case) is $\sim$0.1 dex. For the other nebulae the effect is much lower.

\subsection{Ionic abundances} \label{sec:ionab}
The calculation of the ionic abundances was restricted to the ions needed to compute O, Fe, and Ni ionic abundances and the associated ICFs. We adopted \te([\ion{N}{ii}]) for the calculation of \ioni{O}{+}/\ioni{H}{+}, \ioni{Fe}{+}/\ioni{H}{+}, \ioni{Fe}{++}/\ioni{H}{+}, and \ioni{Ni}{++}/\ioni{H}{+}, and \te([\ion{O}{iii}]) for \ioni{He}{+}/\ioni{H}{+}, \ioni{He}{++}/\ioni{H}{+}, and \ioni{O}{++}/\ioni{H}{+}. 

  \begin{table*}
   \centering
    \caption{Ionic abundances in units of $12+\rm{log}(X^{+i}/H^+)$ and degree of ionization given by O$^{++}$/(O$^{+}$+O$^{++}$).}
    \label{ionabun}
    \begin{tabular}{lr@{}lr@{}lr@{}lr@{}lcr@{}lr@{}lr@{}l}
    \hline  
     &  \multicolumn{2}{c}{\ioni{He}{+}/\ioni{H}{+}} & \multicolumn{2}{c}{\ioni{He}{++}/\ioni{H}{+}} & \multicolumn{2}{c}{\ioni{O}{+}/\ioni{H}{+}} & \multicolumn{2}{c}{\ioni{O}{++}/\ioni{H}{+}}
     & \multicolumn{1}{c}{\ioni{Fe}{+}/\ioni{H}{+}} & \multicolumn{2}{c}{\ioni{Fe}{++}/\ioni{H}{+}$^{a}$} & \multicolumn{2}{c}{\ioni{Ni}{++}/\ioni{H}{+}$^{a}$} & \multicolumn{2}{c}{O$^{++}$/(O$^{+}$+O$^{++}$)}\\     
     \hline
     {\bf Planetary Nebulae} & \multicolumn{15}{c}{}\\
     Cn~1-5      & 11.16 & $\pm$0.02 & $\ldots$   &              & 8.20 & $^{+0.11}_{-0.08}$  & 8.68 & $\pm$0.04 & 5.45:  & 5.99 & $\pm0.06$ (9) & 4.72 & $\pm0.12$ (2) & 0.76 & $^{+0.04}_{-0.07}$\\
     He~2-86    & 11.12 & $\pm$0.02 & $\ldots$   &              & 7.42 & $^{+0.20}_{-0.16}$               & 8.75 & $\pm$0.04 & 4.69:  & 5.42 & $\pm0.09$ (9) & 4.18 & $\pm0.10$ (1) & 0.96 & $^{+0.01}_{-0.02}$\\
     IC~418      & 10.96 & $\pm$0.02 & $\ldots$   &              & 8.45 & $^{+0.32}_{-0.15}$               & 8.09 & $\pm$0.05 & 3.89:  & 4.30 & $^{+0.13}_{-0.08}$ (8) & 2.76 & $^{+0.09}_{-0.04}$: (1) & 0.33 & $^{+0.08}_{-0.12}$\\
     M~1-30      & 11.15 & $\pm$0.02 & $\ldots$   &              & 8.47 & $^{+0.20}_{-0.09}$               & 8.50 & $\pm$0.06 & 4:67:  & 5.52 & $^{+0.12}_{-0.07}$ (6) & 4.21 & $\pm0.10$ (1) & 0.54 & $^{+0.06}_{-0.12}$\\
     M~1-32      & 11.11 & $\pm$0.02  & 8.57 & $\pm$0.07   & 8.22 & $^{+0.29}_{-0.20}$               & 8.26 & $\pm$0.05 & 5.99:  & 6.51 & $\pm0.11$ (9) & 5.18 & $^{+0.13}_{-0.17}$ (1) & 0.52 & $^{+0.11}_{-0.14}$\\
     Mz~3 (P2)  & 10.77 & $\pm$0.02 & $\ldots$   &              & 8.51 & $\pm$0.10              & 6.43 & $\pm$0.04  & 6.01:  & 6.67 & $\pm0.05$ (9) & 5.51 & $\pm0.08$ (2) & 0.01 & $^{+0.01}_{-0.01}$\\ 
     NGC~7009 & 10.97 & $\pm$0.01 & 10.10 & $\pm$0.01 & 6.78 & $\pm$0.04              & 8.62 & $\pm$0.03 & 3.92:  & 4.59 & $\pm0.05$ (8) & 3.86 & $\pm0.07$ (1) & 0.99 & $^{+0.01}_{-0.01}$\\
     NGC~7027 & 10.34 & $\pm$0.02 & 8.61 & $\pm$0.02   & 7.42 & $^{+0.16}_{-0.26}$              & 8.47 & $\pm$0.05 & 4.45:  & 4.56 & $^{+0.08}_{-0.11}$ (8) & 3.26 & $^{+0.11}_{-0.22}$: (1) & 0.91 & $^{+0.04}_{-0.03}$\\
     {\bf \ion{H}{ii} regions} & \multicolumn{15}{c}{}\\
     M8                         & 10.84 & $\pm$0.01 & $\ldots$  & & 8.40 & $^{+0.11}_{-0.03}$ & 7.88 & $\pm$0.02 & 4.64:  & 5.68 & $\pm$0.03 (12) &  4.42 & $^{+0.15}_{-0.18}$ (1) & 0.24 & $^{+0.02}_{-0.05}$\\
     NGC~3576            & 10.94 & $\pm$0.01 & $\ldots$  & & 8.12 & $\pm$0.07 & 8.36 & $\pm$0.04 & 4.59:  & 5.58 & $\pm$0.05 (13) &  4.24 & $^{+0.18}_{-0.28}$ (1) & 0.64 & $\pm0.04$\\     
     Orion (Bar)            & 10.93 & $\pm$0.01 & $\ldots$  & & 8.36 & $^{+0.08}_{-0.05}$ & 8.05 & $\pm$0.02 & 5.19:  & 5.86 & $\pm$0.04 (19) &  4.43 & $\pm$0.04 (5) & 0.33 & $\pm0.04$\\         
     Orion (HH~202 N) & 10.94 & $\pm$0.02 & $\ldots$  & & 8.00 & $^{+0.20}_{-0.10}$ & 8.36 & $\pm$0.06 & 4.70:  & 5.74 & $\pm$0.07 (11) &  4.48 & $\pm$0.08 (1) & 0.71 & $^{+0.04}_{-0.10}$\\
     Orion (HH~202 S) & 10.93 & $\pm$0.02 & $\ldots$  & & 8.23 & $^{+0.18}_{-0.45}$ & 8.10 & $\pm$0.06  & 5.96: &  6.70 & $^{+0.08}_{-0.17}$ (24) &  5.48 & $^{+0.08}_{-0.13}$ (4) & 0.34 & $^{+0.23}_{-0.09}$\\
     Orion (P1)             & 10.96 & $\pm$0.01 & $\ldots$  & & 7.85 & $^{+0.08}_{-0.05}$ & 8.41 & $\pm$0.03 & 4.65:  & 5.51 & $\pm$0.04 (14) &  4.29 & $\pm$0.08 (3) & 0.78 & $\pm0.03$\\     
     Orion (P2)             & 10.98 & $\pm$0.01 & $\ldots$  & & 7.74 & $^{+0.09}_{-0.04}$ & 8.43 & $\pm$0.01 & 4.49:  & 5.33 & $\pm$0.04 (15) &  4.18 & $\pm$0.07 (5) & 0.83 & $\pm0.03$\\
     \hline
    \end{tabular}
    \begin{description}
\item $^a$ We show within parentheses the number of lines used to calculate the ionic abundances.
\end{description}  
  \end{table*}
  
The \ioni{He}{+} abundances were estimated from the emission lines \ion{He}{i} $\lambda$4471, \ion{He}{i} $\lambda$5876, and \ion{He}{i} $\lambda$6678, combining the effective recombination coefficients of \cite{storeyhummer95} for \ion{H}{i} and those of \cite{porteretal12, porteretal13} for \ion{He}{i} that include corrections for collisional excitation and self-absorption effects. The final adopted \ioni{He}{+} abundance is the weighted average (1:3:1; according to the intrinsic intensity ratios of the three lines) of the abundances obtained with the three lines. 
The \ioni{He}{++} abundances were calculated in a few PNe where the \ion{He}{ii} $\lambda$4686 line was detected using the effective recombination coefficients of \cite{storeyhummer95}. 
The \ioni{O}{+} and \ioni{O}{++} abundances were determined from the bright collisionally excited lines: [\ion{O}{ii}] $\lambda\lambda$3726, 3729 and [\ion{O}{iii}] $\lambda\lambda$4959, 5007. 

We have estimated, using the photoionization models from \citet{del14a}, that the contribution of \ioni{Fe}{+} to the total abundance of iron can be more than 20 per cent in the nebulae with O$^{++}$/(O$^{+}$+O$^{++}$)$\lesssim0.7$ and thus, this ion should be taken into account. The calculation of Fe$^{+}$/H$^{+}$ is hampered by the fact that many of the [\ion{Fe}{ii}] lines are affected by fluorescence \citep{rod99, ver00}, though fortunately there is one line almost insensitive to this effect that can be used, [\ion{Fe}{ii}] $\lambda8616$. The line [\ion{Fe}{ii}] $\lambda7155$ can also be used, if the [\ion{Fe}{ii}] $\lambda8616$ is not available, considering that $I$([\ion{Fe}{ii}] $\lambda7155$)/$I$([\ion{Fe}{ii}] $\lambda8616$) $\sim$ 1 \citep{rod99}. We used the emissivities from \cite{bau96} to estimate the values of Fe$^{+}$/H$^{+}$ from these two lines. 

As for the \ioni{Fe}{++} abundances, we used the detections of all the available [\ion{Fe}{iii}] lines from the original references with a few exceptions. Table~\ref{selecFe} shows the [\ion{Fe}{iii}] lines selected to determine the Fe$^{++}$ abundances in the PNe and the \ion{H}{ii} regions studied here. This sample of ionized nebulae presents a very rich spectrum of [\ion{Fe}{iii}] emission lines: between 6 and 24 lines were considered per region (see Table~\ref{ionabun}). A few lines were excluded since their reported fluxes systematically lead to higher abundances than the rest of the lines. We interpret these lines as possible misidentification or blends with nearby lines. The [\ion{Fe}{iii}] $\lambda$4607.03 line was excluded in all the objects because it is strongly blended with the \ion{N}{ii} $\lambda$4607.16 line of multiplet 5. Our analysis shows that the [\ion{Fe}{iii}] $\lambda$4667.01 line returns high abundance values compared to the average value obtained from all the lines in most cases. \cite{garciarojasetal04} reported that this line was blended with an unknown feature in the NGC\,3576 spectra since it presented a wider profile than other [\ion{Fe}{iii}] lines. From a re-inspection of the original spectra of Orion, we also find evidence of a blend with other feature in the blue wing of the line. Therefore, we discarded this line from the sample. The [\ion{Fe}{iii}] $\lambda$8728.84 line was also discarded, because it is contaminated in the low ionization objects by the faint \ion{N}{i} $\lambda$8728.90 line and possibly blended with telluric emission in some objects. And finally, the [\ion{Fe}{iii}] $\lambda$4924.66 line was also excluded in all the objects given that it is potentially affected by the brightest line of multiplet 28 of \ion{O}{ii} at 4924.5 \AA. 

  \begin{table}
   \centering
   \begin{minipage}{50mm}
    \caption{Selected [\ion{Fe}{iii}] lines.}
    \label{selecFe}
    \begin{tabular}{lccc}
    \hline  
     Line & Mult. & \ion{H}{ii} regions &PNe  \\     
     \hline
     3240 &  6F & $\checkmark$ & $\times$ \\
     3286 &  6F & $\checkmark$ & $\times$ \\
     3319 &  6F & $\checkmark$ & $\times$ \\
     3335 &  6F & $\checkmark$ & $\times$ \\
     3356 &  6F & $\checkmark$ & $\times$ \\
     3357 &  6F & $\checkmark$ & $\times$ \\
     3366 &  6F & $\checkmark$ & $\times$ \\
     4008 &  4F & $\checkmark$ & $\checkmark$ \\
     4046 &  4F & $\checkmark$ & $\checkmark$ \\     
     4080 &  4F & $\checkmark$ & $\checkmark$ \\     
     4097 &  4F & $\checkmark$ & $\times$ \\
     4658 &  3F & $\checkmark$ &  $\checkmark$ \\
     4702 &  3F & $\checkmark$ &  $\checkmark$ \\
     4734 &  3F & $\checkmark$ &  $\checkmark$ \\
     4755 &  3F & $\checkmark$ &  $\checkmark$ \\
     4770 &  3F & $\checkmark$ &  $\checkmark$ \\
     4778 &  3F & $\checkmark$ &  $\checkmark$ \\
     4881 &  2F & $\checkmark$ &  $\checkmark$ \\
     4986 &  2F & $\checkmark$ & $\times$ \\
     4990 &  2F & $\checkmark$ & $\times$ \\
     5011 &  1F & $\checkmark$ & $\times$ \\
     5085 &  1F & $\checkmark$ & $\times$ \\
     5270 &  1F & $\checkmark$ &  $\checkmark$ \\
     5412 &  1F & $\checkmark$ & $\checkmark$ \\ 
     6096 &  1F & $\times$ & $\checkmark$ \\ 
     7088 &  1F & $\times$ & $\checkmark$ \\  
     8838 &  8F & $\checkmark$ & $\times$ \\
     9701 &  11F & $\checkmark$ & $\times$ \\
     9960 &  8F & $\checkmark$ & $\times$ \\ \hline
    \end{tabular}
   \end{minipage}
  \end{table}

\begin{figure}
\centering
\includegraphics[width=\hsize, trim = 20 20 50 10, clip =yes]{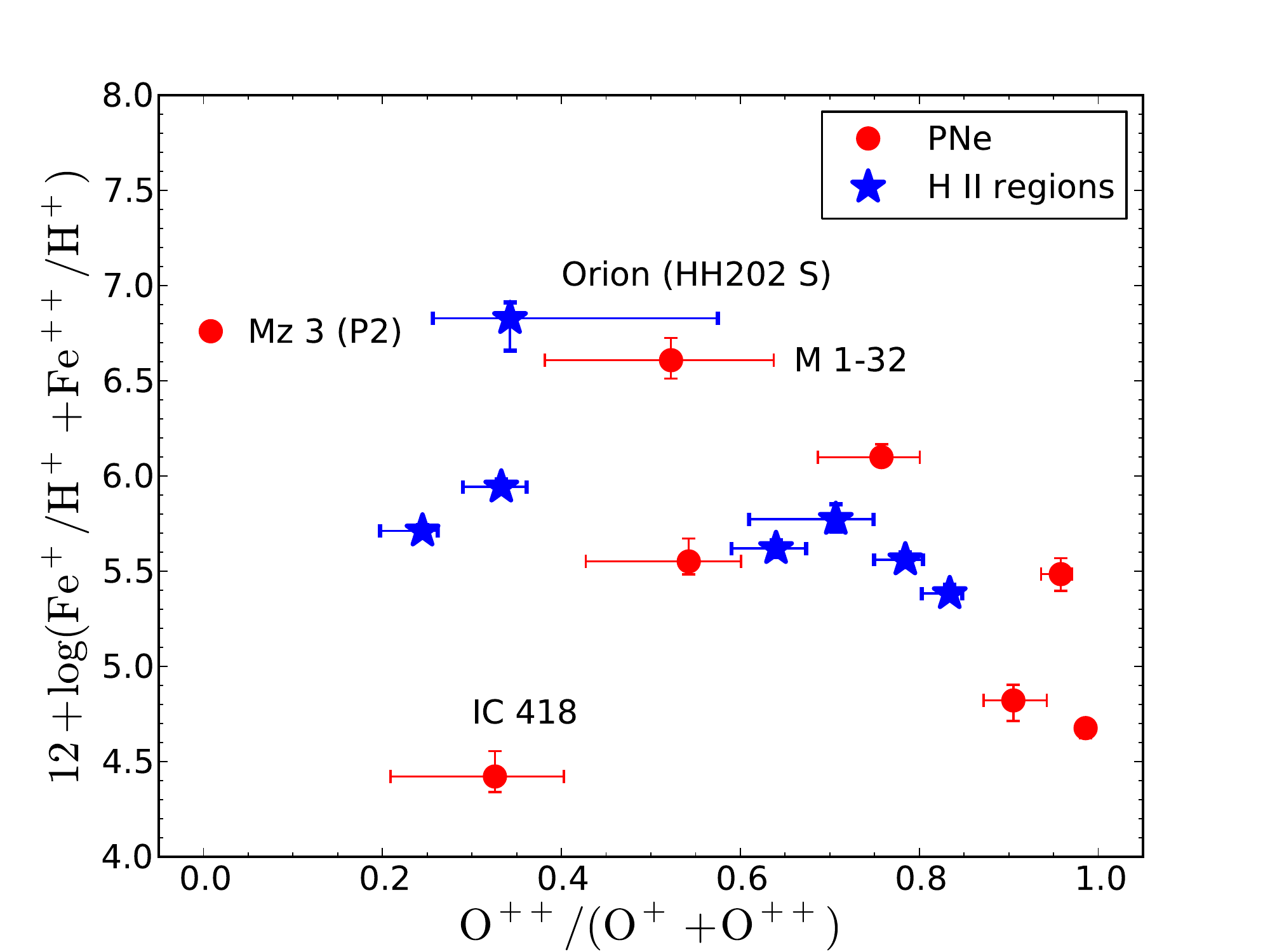}
\includegraphics[width=\hsize, trim = 20 20 50 10, clip =yes]{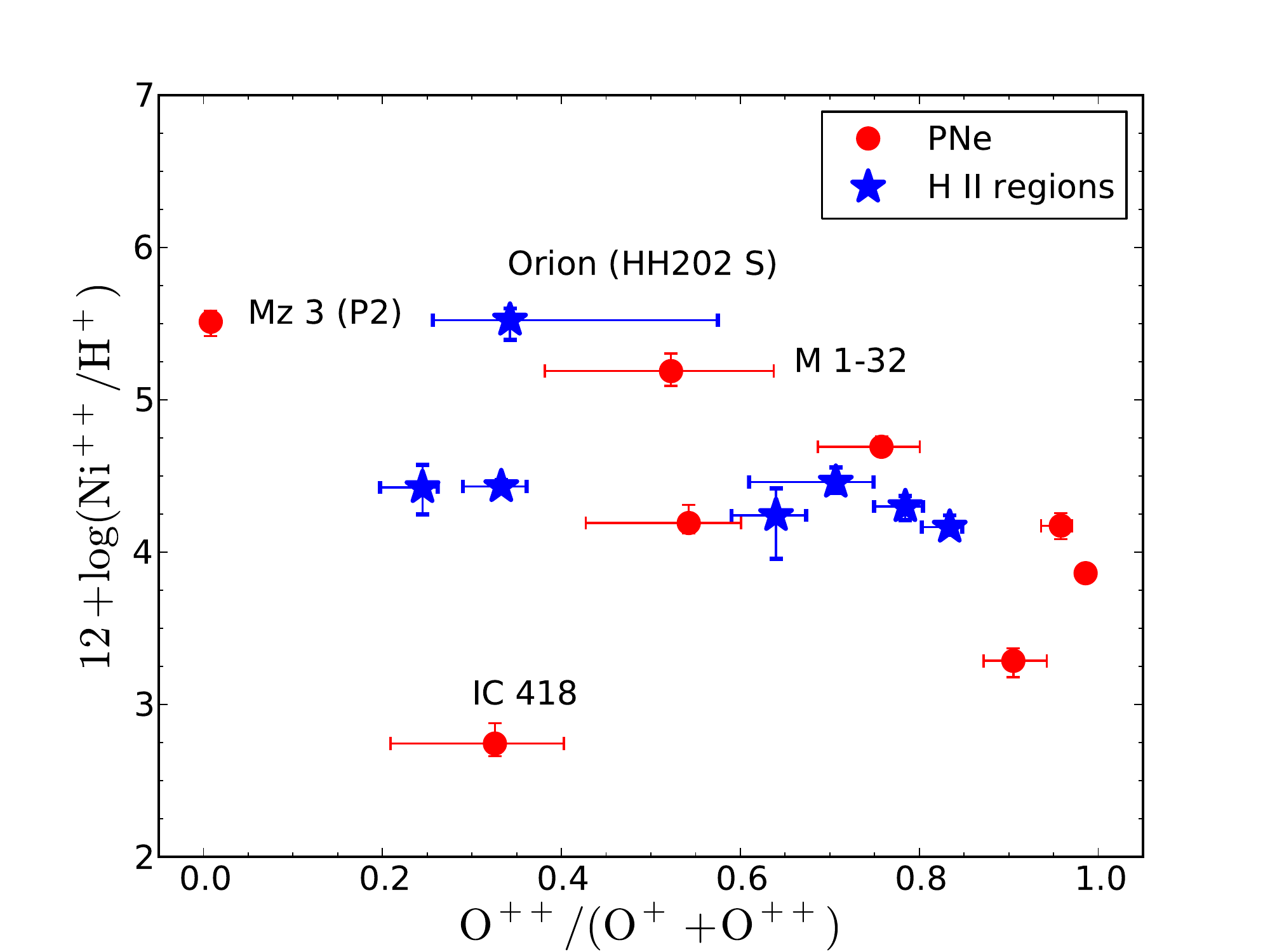}
\caption{Values of \ioni{Fe}{+}/\ioni{H}{+} + \ioni{Fe}{++}/\ioni{H}{+} (upper panel) and \ioni{Ni}{++}/\ioni{H}{+} (lower panel) as a function of the degree of ionization given by O$^{++}$/(O$^{+}$+O$^{++}$) for all the nebulae.\label{fig:sumgi}}
\end{figure}

The values of $\log$(Fe$^{+}$/Fe$^{++}$) in the sample range from $-1.04$~dex (for Orion HH~202 N) to $-0.10$~dex (for NGC~7027). The upper panel of Fig.~\ref{fig:sumgi} shows the values of \ioni{Fe}{+}/\ioni{H}{+} + \ioni{Fe}{++}/\ioni{H}{+} as a function of the degree of ionization for all the PNe and \ion{H}{ii} studied here. There is a clear trend of decreasing \ioni{Fe}{+}/\ioni{H}{+} + \ioni{Fe}{++}/\ioni{H}{+} as the degree of ionization increases, reflecting that for higher ionization objects the presence of highly ionization states becomes significant.

The Ni$^{++}$ abundances were calculated using the lines listed in Table~\ref{selecNi}. We excluded the [\ion{Ni}{iii}] $\lambda$8499.6 line from the analysis because it can be blended with [\ion{Cl}{iii}] $\lambda$8499.7 (in the original references, the feature is labelled either as [\ion{Ni}{iii}] or [\ion{Cl}{iii}]). We also excluded [\ion{Ni}{iii}] $\lambda$7890 in some objects where this emission line was clearly contaminated by night-sky emission. The largest number of [\ion{Ni}{iii}] detections was found in the Orion nebula (see Table~\ref{ionabun}). The analysis in the rest of the sample was based on single line detections and, therefore, these determinations should be regarded with caution since the lines could be misidentified or could be blended with other nearby lines. The Ni$^{++}$ abundances derived for IC~418, M~1-30, NGC~7027 and NGC~7009 rely on the [\ion{Ni}{iii}] $\lambda$7890 line, and M~1-30 is the only PN in which we could inspect the original spectrum and confirm the pure emission of [\ion{Ni}{iii}]. For the other nebulae we place two colons after the Ni$^{++}$ abundances in Table~\ref{selecNi} to denote that these abundances are uncertain. The abundance results of NGC~3576, He~2-86 and M~1-32 also depend on single detections of the line [\ion{Ni}{iii}] $\lambda$6000, while the only reliable feature detected in M8 was [\ion{Ni}{iii}] $\lambda$6401. 

The lower panel of Fig.~\ref{fig:sumgi} shows the values of \ioni{Ni}{++}/\ioni{H}{+} as a function of the degree of ionization for all the PNe and \ion{H}{ii} regions studied here. We observe the same trend than in the upper panel, a clear decrease of the \ioni{Ni}{++}/\ioni{H}{+} ratio as the degree of ionization increases. 

  \begin{table}
   \centering
   \begin{minipage}{50mm}
    \caption{Selected [\ion{Ni}{iii}] lines.}
    \label{selecNi}
    \begin{tabular}{lccc}
    \hline  
     Line & Mult. & \ion{H}{ii} regions &PNe  \\     
     \hline
     4326 &  $^2$D$-^4$P & $\times$ & $\checkmark$ \\
     4597 &  $^3$F$-^1$G& $\checkmark$ &  $\checkmark$ \\
     6000 &  2F & $\checkmark$ &  $\checkmark$ \\
     6401 &  $^3$F$-^3$P& $\checkmark$ &  $\checkmark$ \\
     6534 &  2F & $\checkmark$ &  $\checkmark$ \\
     6946 &  2F & $\checkmark$ &  $\checkmark$ \\
     7890 &  1F & $\checkmark$ &  $\checkmark$ \\  \hline
    \end{tabular}
   \end{minipage}
  \end{table}

\subsection{Total abundances.} \label{sec:totab}
The total oxygen abundances for the \ion{H}{ii} regions were calculated by adding up the ionic abundances of O$^{+}$ and O$^{++}$. As for the PNe, the contribution of higher ionization stages could be significant and we obtain the total abundances of oxygen using the ICF derived in \citet{del14a}. In general, the oxygen abundances derived here are consistent within the uncertainties with the ones obtained in the original papers. The highest differences are of 0.2 dex and they are due to differences in the adopted physical conditions. 

The total iron abundances were derived using the ICF computed by \citet{rodriguezrubin05} from photoionization models:
\begin{equation}\label{icffe1}
\frac{{\rm Fe}}{{\rm O}} = 0.9\left(\frac{{\rm O}^{+}}{{\rm O}^{++}}\right)^{0.08} \frac{{\rm Fe}^{++}}{{\rm O}{^+}}.
\end{equation}
These authors also derived another ICF using observational data (i.e., from nebulae where [\ion{Fe}{iii}] and [\ion{Fe}{iv}] lines have been detected):
\begin{equation}\label{icffe2}
\frac{{\rm Fe}}{{\rm O}} = 1.1\left(\frac{{\rm O}^{+}}{{\rm O}^{++}}\right)^{0.58} \frac{{\rm Fe}^{++}}{{\rm O}{^+}}
\end{equation}
when $\log$(O$^{+}$/(O$^{++}$)$<-0.1$ and 
\begin{equation}\label{icffe3}
\frac{{\rm Fe}}{{\rm O}} = \frac{{\rm Fe}^{+}+{\rm Fe}^{++}}{{\rm O}^{+}}
\end{equation}
otherwise. Note that both ICFs are needed to constrain the real iron abundance. Indeed, \citet{del14b} derived the iron abundances in a group of PNe by adding up the ionic abundances of the main iron ionization states, and they found that the abundances derived in this way are generally intermediate between the two values derived by the ICFs but can be close to either one of them. Therefore, although we do not present here the results, we have checked that our conclusions do not depend on which of the two ICFs for iron is adopted. 

We find a general agreement, within the uncertainties, between our values of Fe/H and the ones computed by other authors. The exception is He~2-86, with a difference of 0.58 dex. This high difference arises from differences in the physical conditions and in the adopted ICF.

\begin{figure}
\centering
\includegraphics[width=\hsize, trim = 10 20 0 10, clip =yes]{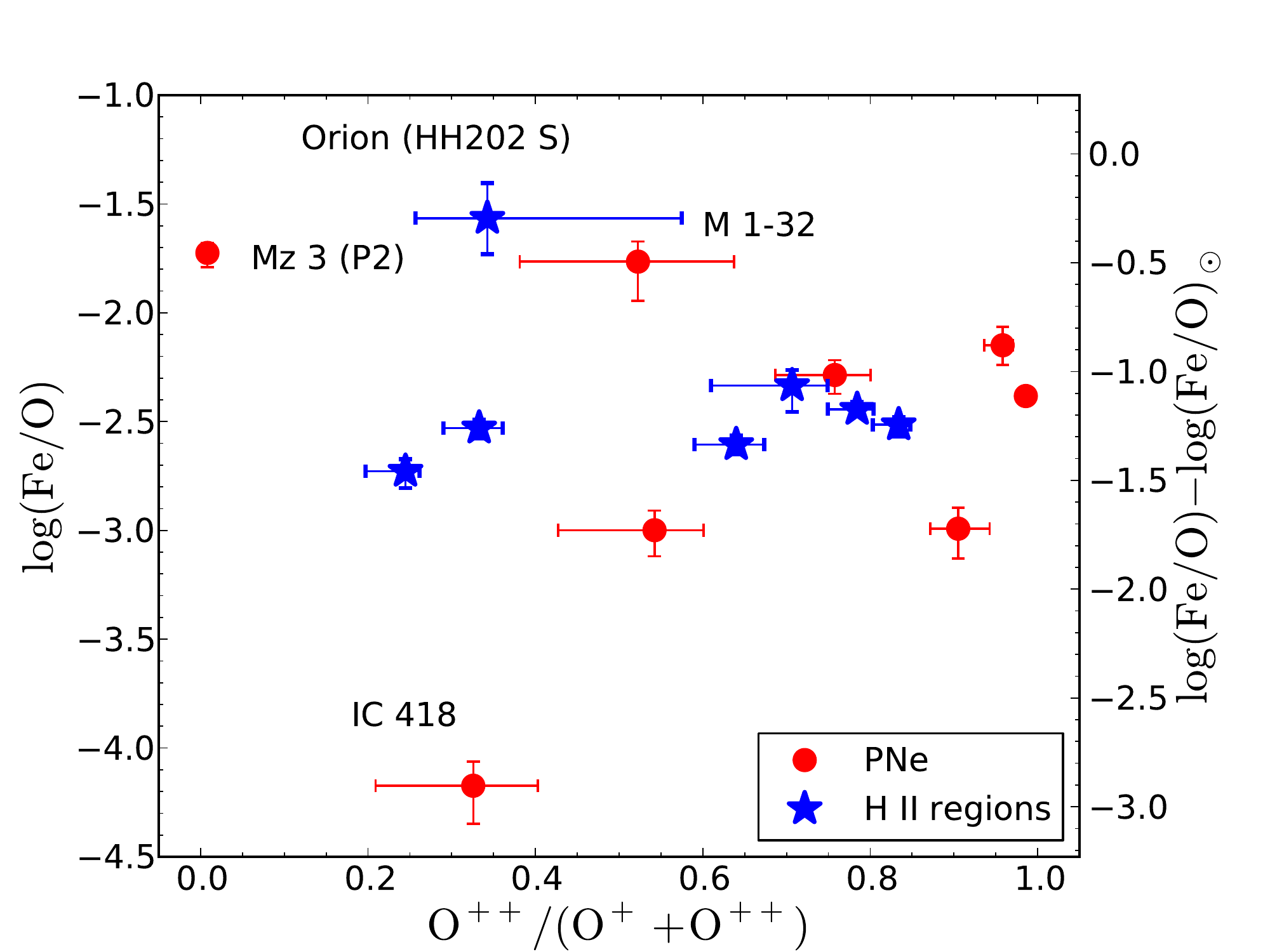}
\includegraphics[width=\hsize, trim = 10 30 0 10, clip =yes]{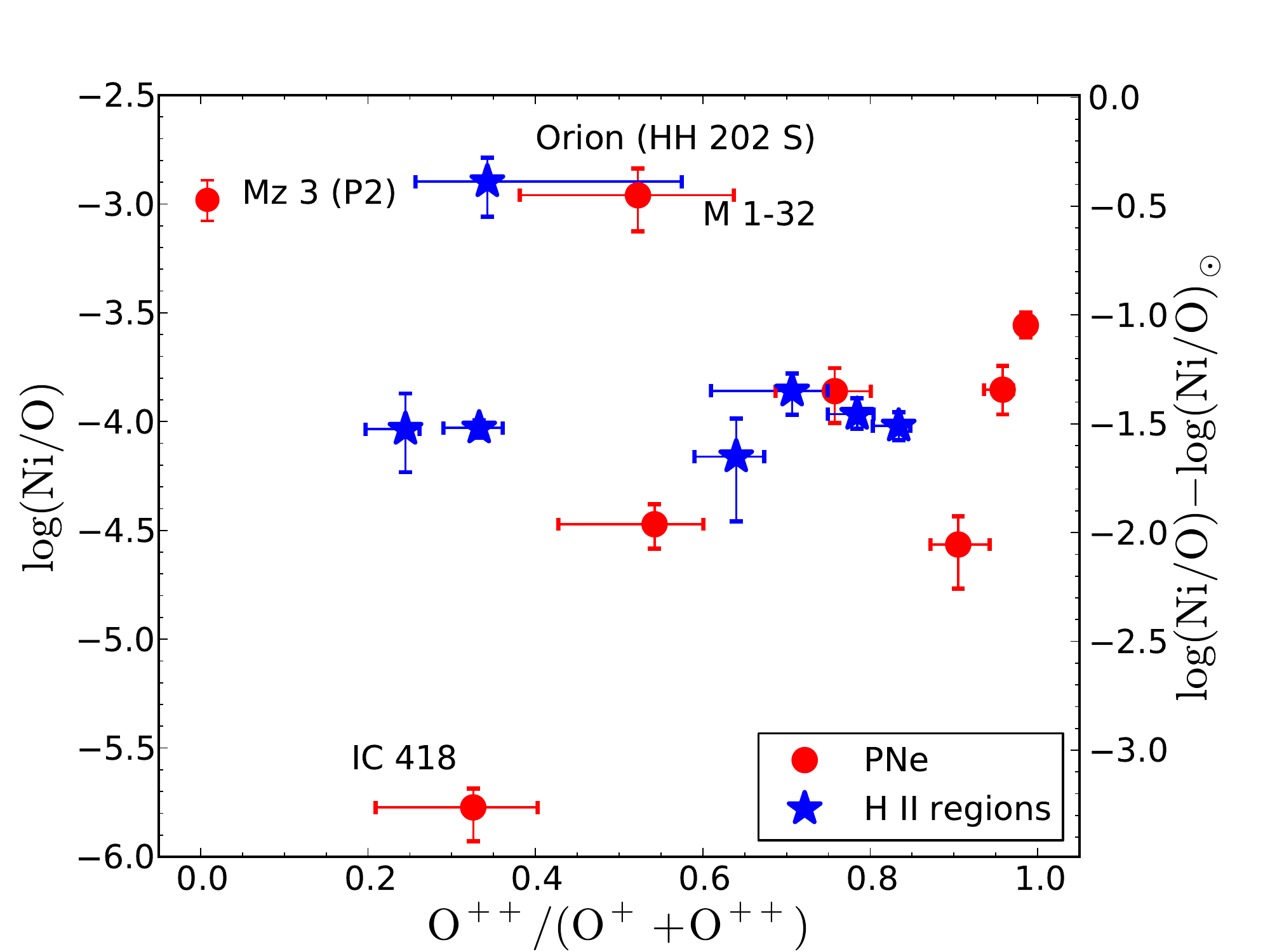}
\caption{Values of Fe/O  and Ni/O (left axis of the upper and lower panels, respectively) and the depletion factors for Fe/O and Ni/O (right axis of the upper and lower panels, respectively) as a function of O$^{++}$/(O$^{+}$+O$^{++}$). The values are derived from the ICF of Eq.~\ref{icffe1} for iron and Eqs.~\ref{icfni1} and \ref{icfni2} for Ni. \label{fig:FeNiOgi}}
\end{figure}

The upper panel of Fig.~\ref{fig:FeNiOgi} shows the Fe/O abundance ratios and the depletion factors for Fe/O (right axis) as a function of the degree of ionization, given by O$^{++}$/(O$^{+}$+O$^{++}$). We adopt $\log$(Fe/O)$_{\odot}$ = $-1.27\pm0.11$ as the reference value \citep{lod10}. The values of Fe/O and [Fe/O] obtained here range from $-4.0$ (IC~418) to $-1.4$ (Orion HH~202~S), and from $-2.69$ to $-0.11$, respectively, in agreement with previous results in the literature \citep[see, e.g,][]{del14a}. Note that the clear separation between IC 418 and the other objects was already present in Fig.~\ref{fig:sumgi}. The PNe cover a wide range of values of Fe/O whereas the \ion{H}{ii} regions all have similar values with the exception of the shocked region in Orion that has a significantly lower iron depletion. 

The absence of a trend in the figure suggests that the adopted ICF is not introducing a bias in our results. 

  \begin{table}
   \centering
   \begin{minipage}{100mm}
    \caption{Total abundances in units of $12+\rm{log}(X/H)$ and $\log$(Fe/Ni).}
    \label{abun}
    \begin{tabular}{lr@{}lr@{}lr@{}lr@{}l}
    \hline  
     & \multicolumn{2}{c}{O/H} & \multicolumn{2}{c}{Fe/H} & \multicolumn{2}{c}{Ni/H} & \multicolumn{2}{c}{Fe/Ni}\\     
     \hline
     \multicolumn{9}{l}{\bf Planetary Nebulae}\\
     Cn~1-5       &  8.81 & $\pm$0.04 & 6.51 & $\pm0.06$ & 4.95 & $\pm0.12$                       & 1.57 & $\pm0.13$\\
     He~2-86     &  8.78 & $\pm$0.05 & 6.63 & $\pm0.08$ & 4.92 & $\pm0.11$                       & 1.70 & $^{+0.09}_{-0.06}$\\
     IC~418       &  8.61 & $^{+0.22}_{-0.13}$ & 4.44 & $^{+0.07}_{-0.10}$ & 2.84 & $\pm0.06$  & 1.60 & $\pm0.07$\\
     M~1-30       & 8.79 & $^{+0.10}_{-0.07}$ & 5.79 & $\pm$0.07 & 4.33 & $\pm0.09$              & 1.46 & $\pm0.11$\\
     M~1-32       & 8.55 & $^{+0.14}_{-0.08}$ & 6.78 & $\pm$0.06 & 5.62 & $^{+0.13}_{-0.16}$ & 1.16 & $^{+0.16}_{-0.12}$\\
     Mz~3 (P2)  & 8.52 & $\pm0.09$ & 6.80 & $\pm$0.05  & 5.55 & $\pm$0.08                           & 1.25 & $\pm0.07$\\ 
     NGC~7009 & 8.66 & $\pm$0.03 & 6.28 & $\pm0.04$ & 5.11 & $\pm0.07$                            & 1.18 & $\pm0.06$\\
     NGC~7027 & 8.52 & $\pm$0.05 & 5.51 & $\pm0.09$ & 3.96 & $^{+0.12}_{-0.20}$               & 1.57 & $^{+0.19}_{-0.12}$\\
     \multicolumn{9}{l}{\bf \ion{H}{ii} regions}\\
     M8               & 8.51 & $^{+0.07}_{-0.04}$  & 5.79 & $\pm$0.03              & 4.50 & $^{+0.14}_{-0.19}$ & 1.29 & $^{+0.18}_{-0.15}$\\
     NGC~3576  & 8.56 & $\pm$0.03   & 5.95 & $\pm$0.04              & 4.41 & $^{+0.17}_{-0.29}$ & 1.53 & $^{+0.33}_{-0.15}$\\    
     Orion (HH~202 N) & 8.52 & $\pm$0.07               & 6.18 & $\pm$0.06              & 4.67 & $\pm0.08$             & 1.50 & $\pm0.08$\\
     Orion (HH~202 S) & 8.47 & $^{+0.23}_{-0.14}$   & 6.93 & $\pm$0.06              & 5.58 & $^{+0.09}_{-0.05}$   & 1.34 & $\pm0.05$\\
     Orion (P1)   & 8.52 & $\pm$0.02   & 6.07 & $\pm$0.03               & 4.55 & $\pm0.07$             & 1.53 & $\pm0.07$\\     
     Orion (P2)   & 8.51 & $\pm$0.02   & 6.00 & $\pm$0.03               & 4.49 & $\pm0.06$              & 1.50 & $\pm0.07$\\
     Orion (Bar)  & 8.54 & $\pm$0.05   & 6.01 & $\pm$0.03              & 4.51 & $\pm0.04$              & 1.50 & $\pm0.04$\\         
     \hline
    \end{tabular}   
   \end{minipage}
  \end{table}

\subsubsection{A new ionization correction factor for nickel
\label{sec:icfs}}

The few determinations of nickel abundances in the literature use two ICFs based on the similarity of ionization potentials: Fe/Ni = Fe$^+$/Ni$^+$ \citep{lucy95, zhangliu02} and Ni/O =  (Ni$^+$+Ni$^{++}$)/O$^+$ \citep{mesadelgadoetal09b}. Using the grid of photoionization models computed in \citet{del14a}, we found that, in the absence of fluorescence effects, the first approach can either underestimate or overestimate $\log$(Fe/Ni) by $\sim$0.3 dex in the nebulae with low degrees of ionization and can be even worse, underestimating $\log$(Fe/Ni) by up to 0.7 dex, in the nebulae with very high degrees of ionization. The second ICF is valid only for nebulae with ionizing stars with low effective temperature (below 50\,000 K) and low degrees of ionization (O$^{++}$/(O$^{+}$+O$^{++}$) < 0.2), otherwise this ICF overestimates the $\log$(Ni/O) ratio by up to 0.7 dex.

We explored the possibility of obtaining an ICF for Fe/Ni based on Fe$^{++}$ and Ni$^{++}$ abundances, but we found that a better approach to derive nickel abundances is to use an ICF based on the Ni$^{++}$ and O$^{+}$ abundances:
\begin{equation}
\frac{{\rm Ni}}{{\rm O}} = \frac{{\rm Ni}^{++}}{{\rm O}^+} {\rm ICF}({\rm Ni}^{++}/{\rm O}{^+}).
\end{equation}
Using the photoionization models from \citet{del14a}, we performed two fits of the $x$(O$^+$)/$x$(Ni$^{++}$) values as a function of O$^{++}$/(O$^{+}$+O$^{++}$). The first one:
\begin{equation}
{\rm ICF}({\rm Ni}^{++}/{\rm O}{^+}) = 1.1 - 0.9\frac{{\rm O}^{++}}{({\rm O}^{+}+{\rm O}^{++})}
\label{icfni1}
\end{equation}
is valid when \ion{He}{ii} lines are not detected. Otherwise, the expression:
\begin{equation}
{\rm ICF}({\rm Ni}^{++}/{\rm O}{^+}) = 2.5 - 2.3\frac{{\rm O}^{++}}{({\rm O}^{+}+{\rm O}^{++})}
\label{icfni2}
\end{equation}
should be used. 

\begin{figure}
\centering
\includegraphics[width=\hsize, trim = 20 0 50 30, clip =yes]{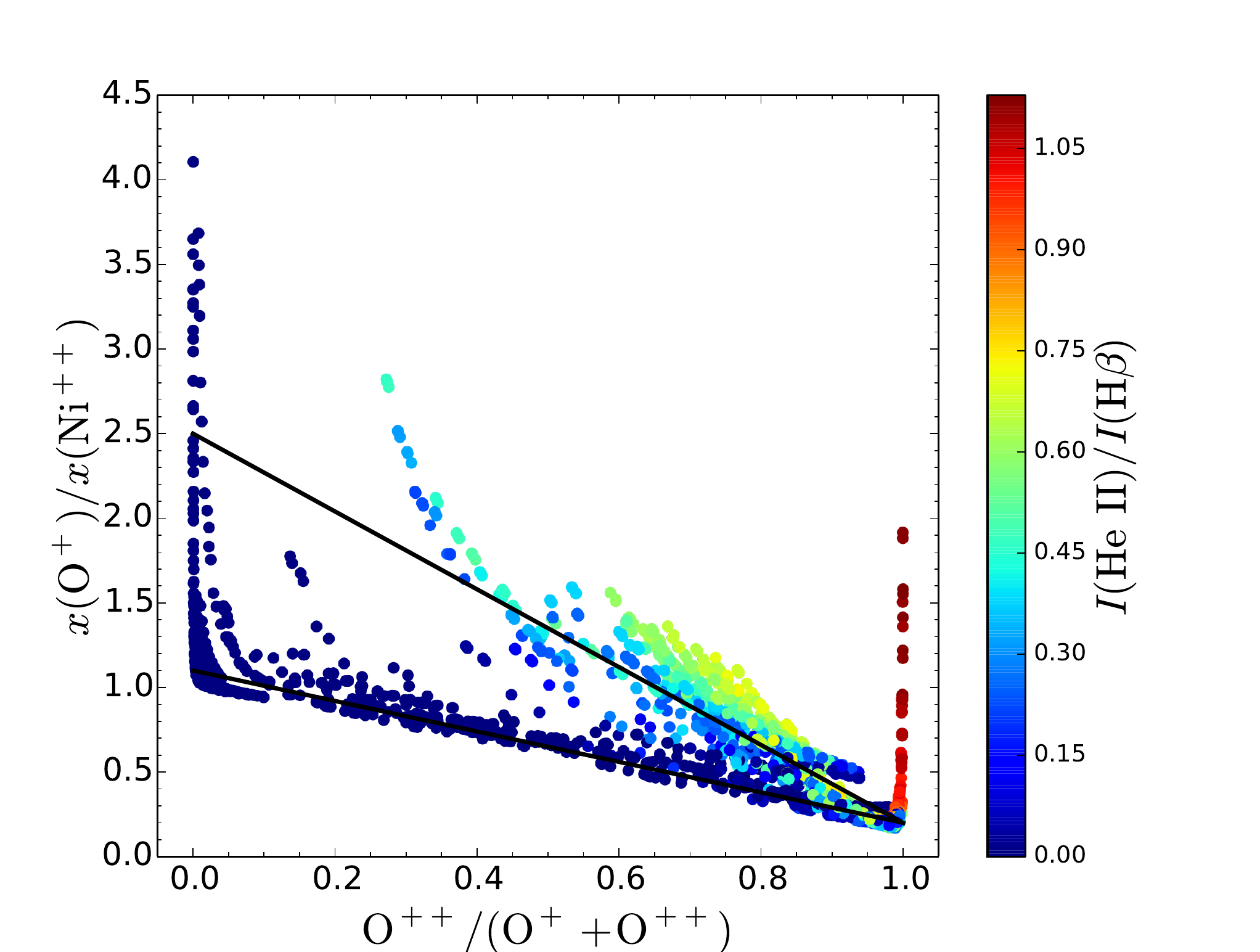}
\includegraphics[width=\hsize, trim = 20 0 50 30, clip =yes]{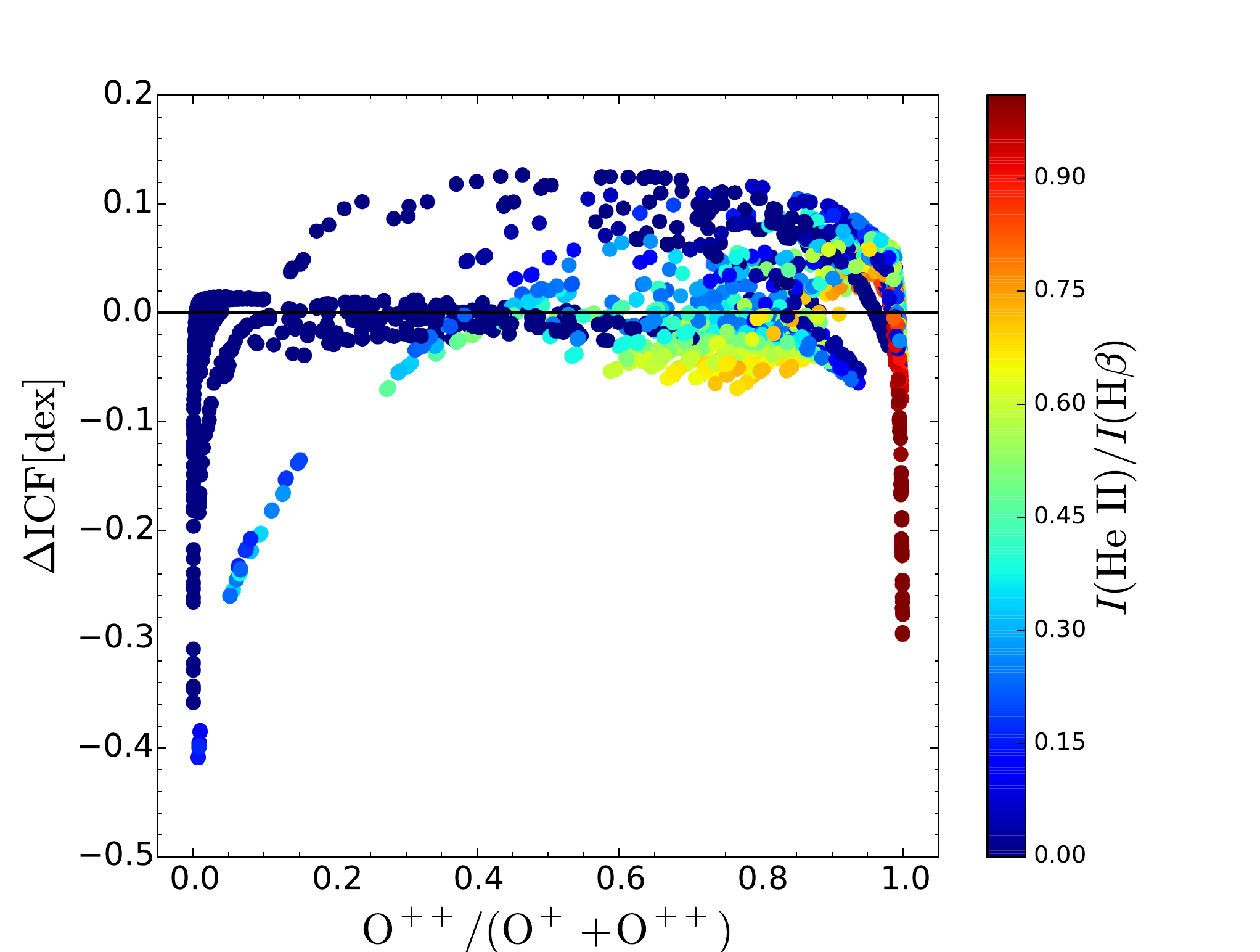}
\caption{{\it Upper panel:} Values of $x$(O$^{+}$)/$x$(Ni$^{++}$) as a function of O$^{++}$/(O$^{+}$+O$^{++}$). The solid lines represent our two fits to the values, that is ICF(Ni$^{++}$/O$^{+}$). {\it Lower panel:} Differences in dex between the values of $x$(O$^{+}$)/$x$(Ni$^{++}$) from the models and the values of ICF(Ni$^{++}$/O$^{+}$) obtained using our fit. The solid line represents where the values of $x$(O$^{+}$)/$x$(Ni$^{++}$) from the models and the ICF we derived from the fit are equal. The colour bar located on the side of both plots runs from low to high values of $I$(\ion{He}{ii})/$I$(H$\beta$).
\label{fig:icfni}}
\end{figure}

The upper panel of Fig.~\ref{fig:icfni} shows the photoionization models together with the two fits, whereas the lower panel displays the uncertainties in the Ni/O ratio associated with the use of the above ICF. The figure shows that the uncertainties in Ni/O are within $\pm$0.1 dex in most cases. Like all the ICFs derived from photoionization models they should be taken with caution since the physics involved in the calculations may be incomplete. Besides, the ICFs are also affected by uncertainties in the adopted atomic data.

Our values of Ni/H and the ones reported in the original papers differ by less than 0.2 dex in most objects. The exceptions are Cn~1-5, Orion (HH~202 N), and He~2-86, with differences of 0.32, 0.36, and 0.63 dex, respectively, which are due to differences in the physical conditions and in the adopted ICF.

The lower panel of Fig.~\ref{fig:FeNiOgi} shows the abundance ratio Ni/O and the depletion factor for Ni/O as a function of the degree of ionization, given by O$^{++}$/(O$^{+}$+O$^{++}$). We adopt $\log$(Ni/O)$_{\odot}$ = $-2.51\pm0.08$ as the reference value \citep{lod10}. As it happens for Fe/O, there is no evident trend between the abundance ratios and the degree of ionization, suggesting that the nickel abundances derived with the ICF proposed here are not biased. The values of $\log$(Ni/O) obtained in the sample range from $-5.61$ to $-2.87$ and the respective depletion factors vary from $-3.11$ to $-0.26$; these results are similar to those found for Fe/O. Again, IC~418 and Orion (HH~202~S) are the objects with the highest and lowest depletions, respectively.  

\begin{figure}
\centering
\includegraphics[width=\hsize, trim = 10 0 10 30, clip =yes]{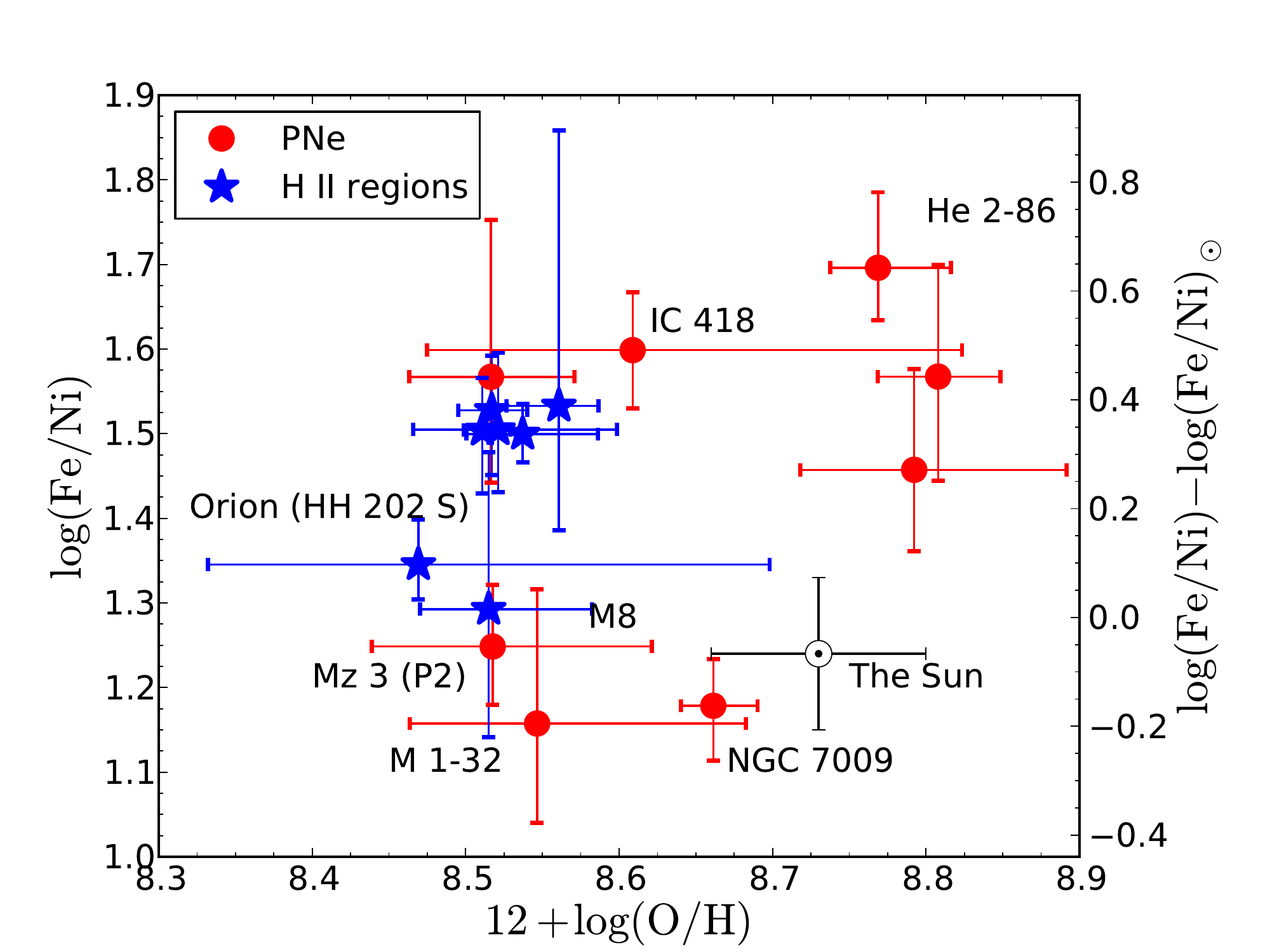}
\caption{Values of Fe/Ni as a function of O/H for the group of PNe (circles) and \ion{H}{ii} regions (stars) studied here.\label{fig:FeNiOH}}
\end{figure}

Fig.~\ref{fig:FeNiOH} shows the values of Fe/Ni and the depletion factor for Fe/Ni as a function of O/H. The objects have oxygen abundances between 12 + $\log$ = $8.47\pm0.11$ (Orion HH~202 S) and $8.81\pm0.06$ (Cn~1-5). We see from the figure that most of the nebulae have higher Fe/Ni ratios than the Sun (by $\sim0.3$ dex) and lower O/H ratios (by $\sim0.2$ dex). We also see that there is no trend between the Fe/Ni ratios and the oxygen abundances. 

\citet{del15} found evidence of oxygen enrichment in some Galactic PNe descending from near-solar metallicity stars. Three of the PNe studied here, IC~418, NGC~7027, and Mz~3, could be oxygen enriched according to their chlorine and argon abundances derived in the manner described in \citet{del15}. In these three PNe oxygen might not be an adequate proxy for the metallicity. However the conclusions described below do not depend on whether we use chlorine, argon, or oxygen as metallicity. Since oxygen is a good metallicity indicator for most of the sample and since the oxygen abundances are the most reliable ones that can be calculated in ionized nebulae, we use oxygen throughout the paper, instead of chlorine or argon.

\citet{zhangliu06} derived $\log$(Fe/Ni) = 0.68 for IC~418 and \citet{mesadelgadoetal09b} obtained 1.07 and 1.08 for the nebular and shock components of Orion (HH~202), respectively. Our derived values of $\log$(Fe/Ni) for these objects are higher by 0.3--1.0 dex. This arises from the differences in the adopted lines and ICFs. As we argue above, our method should provide more reliable Fe/Ni ratios. 

\section{Results and discussion}
\label{discussion}

\begin{figure}
\centering
\includegraphics[width=\hsize, trim = 20 0 20 0, clip =yes]{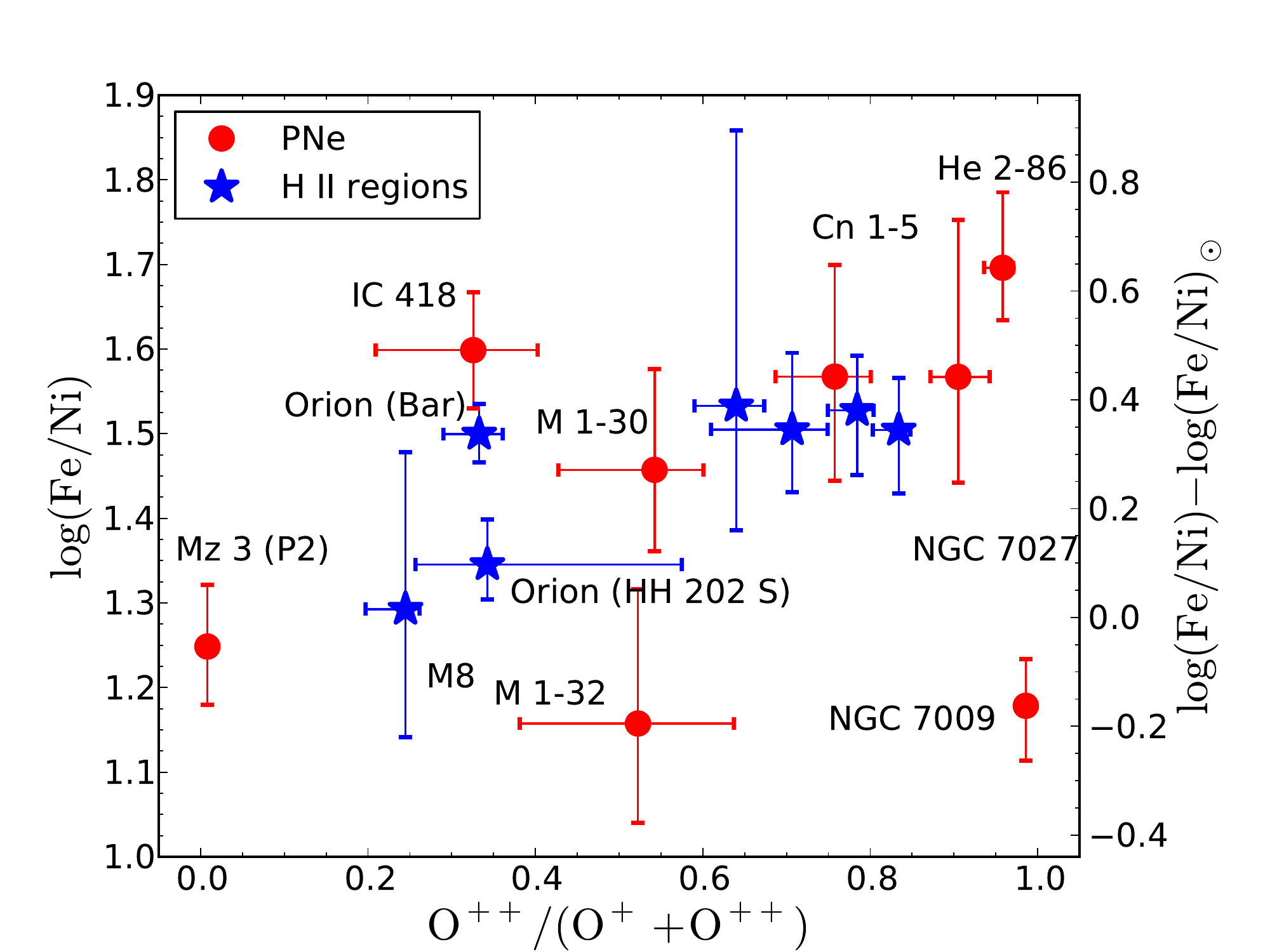}
\caption{Values of Fe/Ni as a function of O$^{++}$/(O$^{+}$+O$^{++}$) for the group of PNe (circles) and \ion{H}{ii} regions (stars) studied here.\label{fig:FeNigi}}
\end{figure}

In Fig.~\ref{fig:FeNigi} we show the Fe/Ni ratios and the depletion factors for Fe/Ni as a function of the degree of ionization. Some of the dispersion in this figure is probably related to the adopted ICFs, but we see that there is no obvious trend between the Fe/Ni ratios and the degree of ionization. Therefore, our correction scheme to compute iron and nickel abundances seems to work correctly. 

It can be seen in this figure that most of the \ion{H}{ii} regions have similar Fe/Ni abundances ratios whereas the PNe cover a wider range of values. This suggests that there are larger variations in the dust formation, growth, and destruction efficiencies in PNe than in \ion{H}{ii} regions, as already found by \citet{delgadoingladaetal09}. 

We have explored whether a change in the determination of the adopted physical conditions would modify our results. To do this, we recalculated the Fe/Ni ratios using three other different approaches: 1) using only \nel[\ion{Cl}{iii}] and two \te\ (the ones derived from [\ion{N}{ii}] and [\ion{O}{iii}] lines), 2) using the average \nel\ implied by the 2--3 available diagnostic ratios from Table~\ref{phycond} and \te[\ion{N}{ii}] for the temperature, and 3) using the average \nel\ and \te[\ion{O}{iii}]. We found minor differences of $\sim$0.1 dex in the final $\log$(Fe/Ni) ratios obtained through the different methods. Therefore, our conclusions do not critically depend on the initial assumptions made about the physical conditions. 

\begin{figure}
\centering
\includegraphics[width=\hsize, trim = 0 0 40 30, clip =yes]{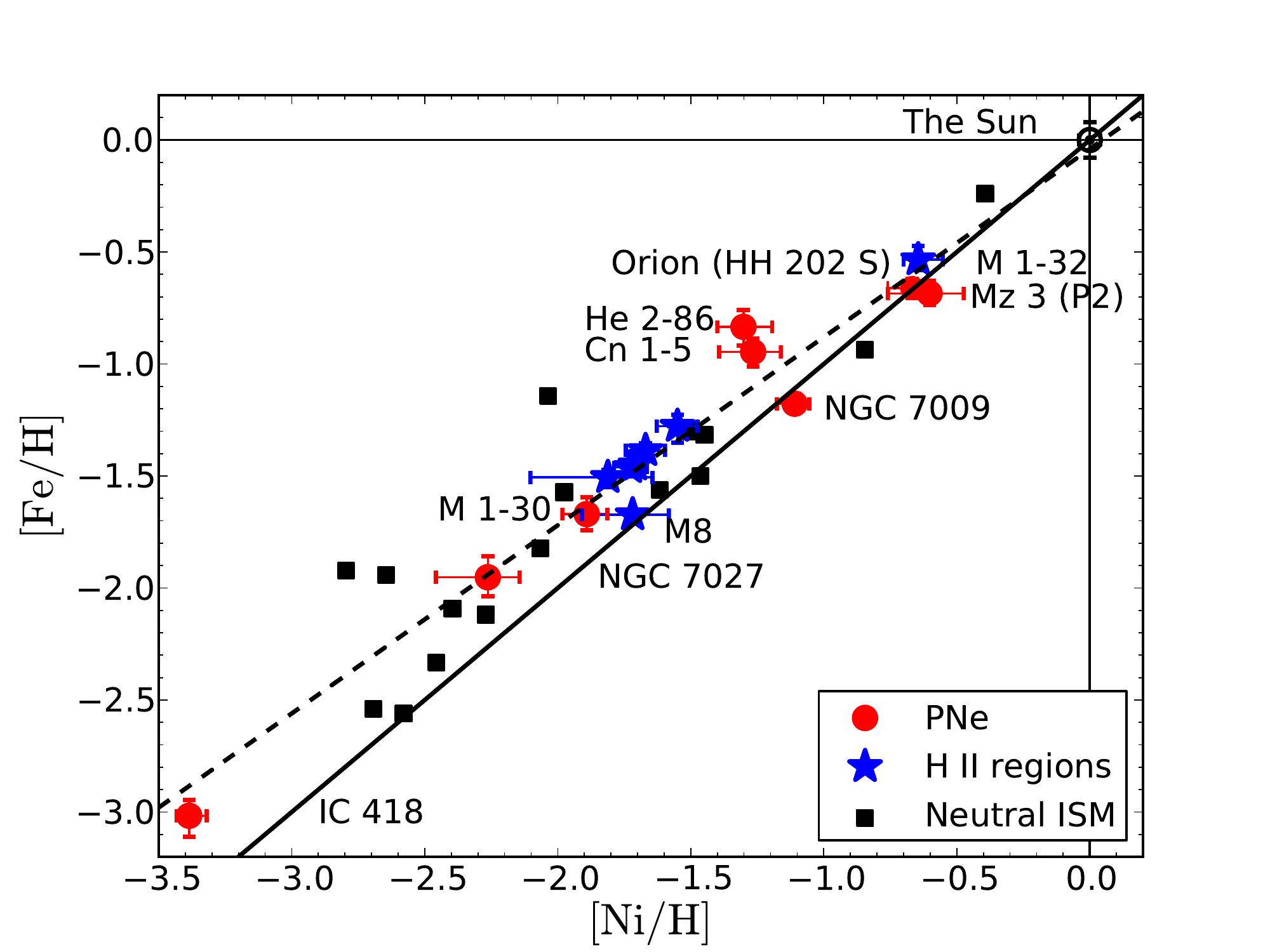}
\caption{The Fe depletion factors as a function of the Ni depletion factors for the group of PNe (circles) and \ion{H}{ii} regions (stars) studied here. The dashed line shows a linear fit to the data: $y = 0.85x -0.03$ whereas the solid line represents equal depletions for Fe and Ni. We include the value for the Sun (solar symbol) and those obtained from \citet{jen09} for 16 neutral clouds in the solar neighbourhood (squares).\label{fig:FeNi}}
\end{figure}

Fig.~\ref{fig:FeNi} shows the comparison between the iron and nickel depletions for all the studied nebulae. There is an evident correlation between both depletion factors. The solid line in the figure indicates where [Fe/H] and [Ni/H] are equal. The dashed line shows a linear fit to the data, $y = 0.85x - 0.03$. The Pearson correlation coefficient between both parameters is $r = 0.98$ with a $p-$value (the probability of finding this result when the correlation coefficient is zero) of $1.9\times10^{-11}$. The fact that most of the objects lie above the solid line suggests that the nickel atoms are somewhat more depleted than the iron ones. Note that although the depletions in the sample cover a wide range of values (of around 3 dex), the trend between [Fe/H] and [Ni/H] is clear in the whole range. This is another indication that our derived abundances of iron and nickel are good estimates. Among the three ionized nebulae with the lowest depletions, Orion HH~202 S, M~1-32, and Mz~3, the first one is a shocked region of Orion where dust destruction might have taken place, the second has high-velocity outflows \citep{garciarojasetal12, rec14} where dust destruction could also have occurred, and the third one probably contains a symbiotic star \citep{cly15} where perhaps dust formation has not happened.  

For comparison, we have also included in the plot the values for the Sun \citep[with 12+$\log$(Fe/H)$_\odot$ = $7.46\pm0.08$ and 12+$\log$(Ni/H)$_\odot$ = $6.22\pm0.04$;][]{lod10} and for the 16 diffuse clouds located in the solar neighbourhood that have available data to calculate their iron and nickel abundances (taken from \citet{jen09} and corrected as this author suggests). The figure reveals that the diffuse clouds show the same trend between the iron and nickel depletions as the ionized nebulae.

It would be better to use oxygen, argon, or chlorine instead of hydrogen in Fig.~\ref{fig:FeNi} because we expect the intrinsic value of iron and nickel relative to these elements to vary less from one object to another than Fe/H and Ni/H. However, the number of neutral clouds in \citet{jen09} with available values of O/H and Cl/H is scarce and there are no available data of the Ar abundances. We have checked that the behaviour is similar if we replace hydrogen with oxygen, argon, and chlorine for the ionized nebulae.

\begin{figure}
\centering
\includegraphics[width=\hsize, trim = 15 0 30 0, clip =yes]{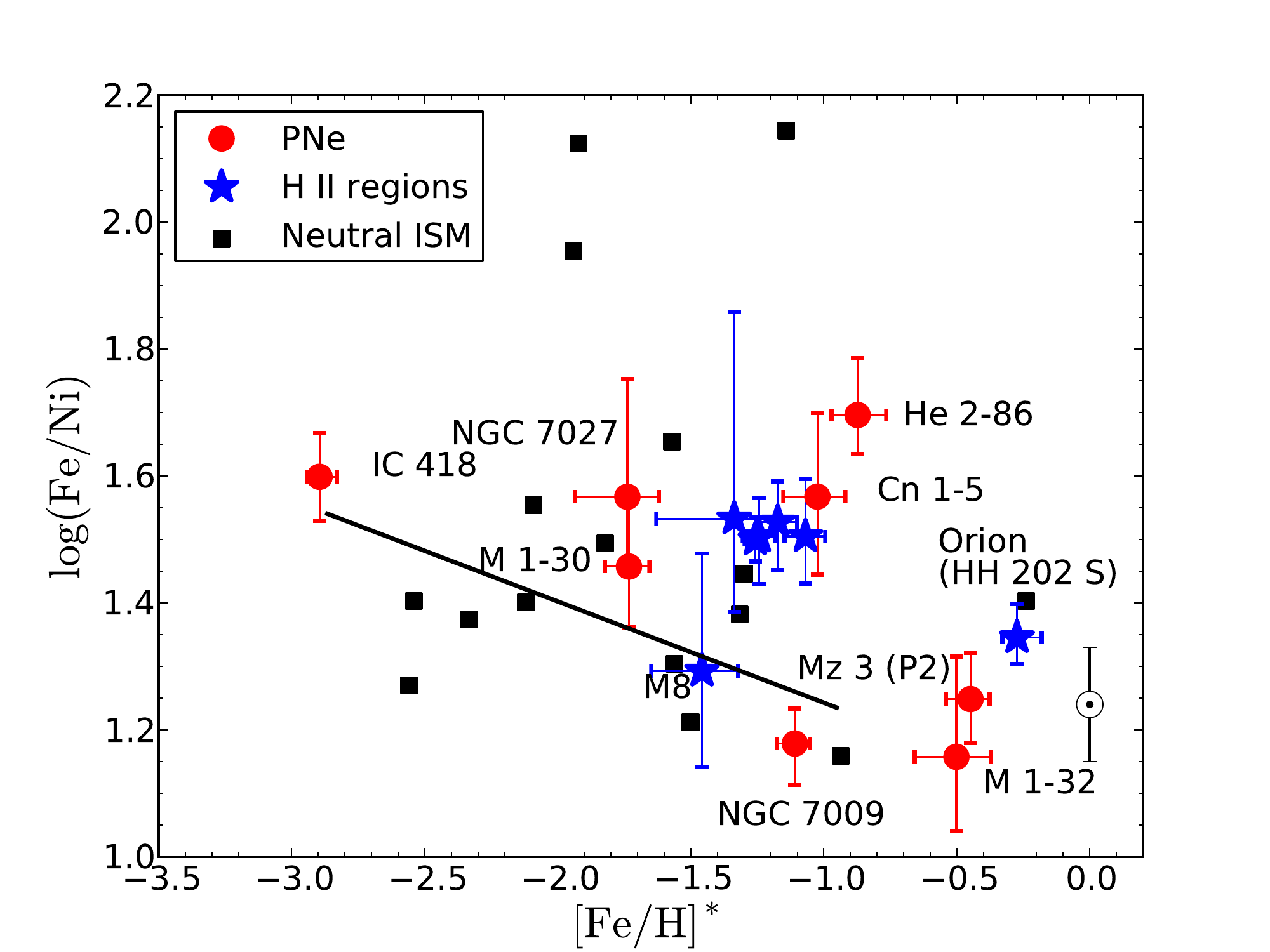}
\includegraphics[width=\hsize, trim = 15 0 30 0, clip =yes]{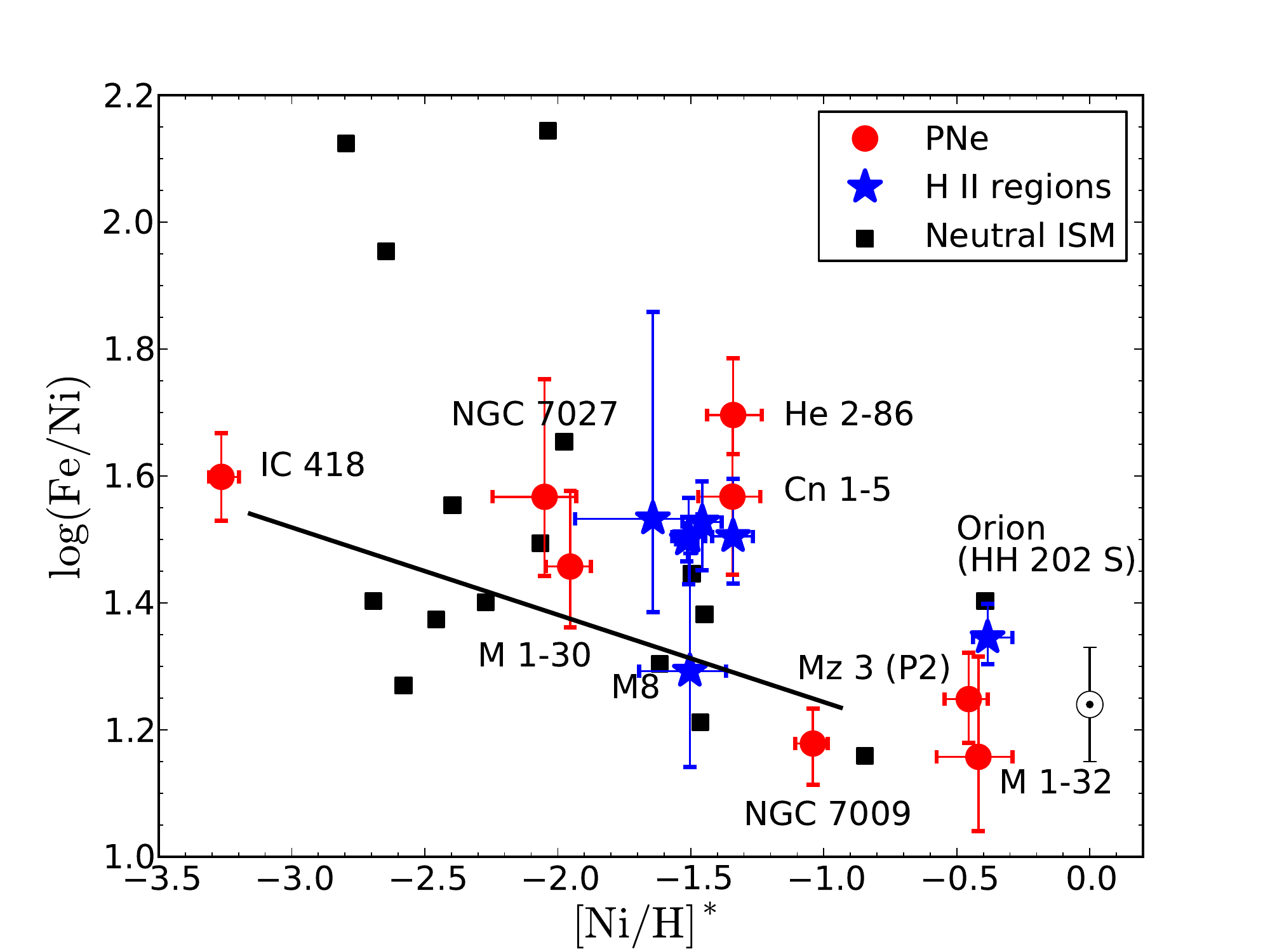}
\caption{Values of Fe/Ni as a function of the depletion factor for Fe (upper panel) and Ni (lower panel) for the group of PNe (circles) and \ion{H}{ii} regions (stars) studied here. The depletion factors are calculated differently from those in Fig.~\ref{fig:FeNi} (see text). We include the value for the Sun (solar symbol) and those obtained from \citet{jen09} for 16 neutral clouds in the solar neighbourhood (squares). The solid lines correspond to the fits provided by \citet{jen09} to characterize Fe and Ni depletions. \label{fig:FeNid}}
\end{figure}

Finally, in Fig.~\ref{fig:FeNid} we show the Fe/Ni abundance ratios as a function of the depletion factor given by [Fe/H] (upper panel) and [Ni/H] (lower panel). The values for the Sun and the neutral clouds from \citet{jen09} are also included in the figure. As shown in Fig.~\ref{fig:FeNiOH}, some PNe have higher O/H ratios than the others, and we decided to take this into account to compute the [Fe/H] and [Ni/H] values that are presented in Fig.~\ref{fig:FeNid}. We name these values as [Fe/H]$^*$ and [Ni/H]$^*$ in the figure to distinguish them from those shown in Fig.~\ref{fig:FeNi} that are somewhat different. We calculated these depletion factors for Fe and Ni using as reference, instead of the solar abundances, the Fe/H and Ni/H obtained from the (Fe/O)$_\odot$ or (Ni/O)$_\odot$ abundance ratios and the O/H derived for each object. We could not use directly the [Fe/O] and [Ni/O] values because most of the neutral clouds do not have available measurements of the oxygen abundance. The solid lines in the figures correspond to the fits provided by \citet{jen09} to describe the depletions of these elements and obtained from observations of neutral clouds of the solar neighbourhood through different lines of sight. The ranges of depletions covered by the solid lines in this figure correspond to the ones used \citet{jen09} in his Fig.~7. 

Fig.~\ref{fig:FeNid} illustrates the clear correlation that exists between the Fe/Ni ratios and the depletion factors. Even with some dispersion, that could be due to the uncertainties associated with the calculations, there is a general agreement between the values obtained in the 16 neutral interstellar clouds, in the 11 ionized nebulae studied here, and the trend derived by \citet{jen09} using a larger sample of neutral clouds. The trend indicates that when the depletion is low, [Fe,Ni/H] $>-1.0$, the Fe/Ni ratios are near solar, whereas for higher depletions, [Fe,Ni/H] $<-1.0$, the Fe/Ni ratios increase as does the depletion. As discussed by \citet{jen09}, this kind of trends reflect how different elements accrete onto dust grains when grains grow in the ISM. The result in Fig.~\ref{fig:FeNid} suggests that nickel atoms have a higher sticking efficiency than iron atoms in regions with higher depletions, that include the neutral and ionized regions of the ISM and also the ionized gas around evolved stars. 

If we had used the other ICF proposed by \citet{rodriguezrubin05} for iron (Eqs.~\ref{icffe2} and \ref{icffe3}), the values of $\log$(Fe/Ni) would be up to 1 dex lower (for NGC~7009) or 0.4 dex higher, and the dispersion of the results would increase, but the general behaviour between the Fe/Ni ratios and the depletion factors remains similar. 

\section{Conclusions} \label{conclu}

We have used deep and high-quality spectrophotometric data from the literature to compute the iron and nickel abundances in eight planetary nebulae (PNe) and three \ion{H}{ii} regions in our Galaxy. We have homogeneously derived the physical conditions, \te\  and \nel, and the ionic abundances for the whole sample, making use of state-of-the-art atomic data. Our working set of Fe and Ni lines has been carefully selected from the literature, avoiding blends with other lines and/or telluric features. 

The total iron abundances were computed using [\ion{Fe}{ii}] and [\ion{Fe}{iii}] lines and the ICFs derived by \citet{rodriguezrubin05}. For nickel, we have avoided using the [\ion{Ni}{ii}] lines because most of them are likely affected by fluorescence. We have only used [\ion{Ni}{iii}] lines and the correction scheme derived here from a large grid of photoionization models. 

We have compared the depletions of iron and nickel in the ionized nebulae with the ones derived in 16 neutral clouds from the solar neighbourhood \citep{jen09} that have available data to calculate the depletions of iron and nickel. We found that both the ionized and the neutral nebulae show a clear correlation between both depletion factors, that remains the same over the whole range of  $\sim3$ dex covered by the depletions. The iron and nickel depletion patterns are similar in the PNe, where the dust is fresh, and in the \ion{H}{ii} regions and the neutral ISM, where the dust is processed dust. We also found a clear correlation between the Fe/Ni ratios and the depletion factors, [Fe/H] and [Ni/H]. Although there is some dispersion, the results obtained for the ionized and diffuse nebulae and the expressions derived by \citet{jen09} to characterize element depletions from data of a large sample of neutral clouds, all broadly agree with: 1) the Fe/Ni abundance ratios are close to the solar value for low depletions, [Fe, Ni/H] $>-1$ and 2) the Fe/Ni ratios increase as the depletion increases for [Fe, Ni/H] $<-1$. These results suggest that nickel atoms are more efficiently incorporated to the dust grains than iron atoms in ambients where the dust formation or growth has been more efficient. 

\section*{Acknowledgements}
The authors thank the referee, Grazyna Stasi\'nska, for her useful comments. GDI gratefully acknowledges support from Fundaci\'on GSI and from the Mexican CONACYT grant CB-2014-241732. AMD acknowledges support from the FONDECYT grant 3140383. JGR acknowledges support from Severo Ochoa excellence program (SEV-2011-0187) postdoctoral fellowship. MR acknowledges support from Mexican CONACYT grant CB-2014-240562. JGR and CE acknowledge support from the Spanish Ministry of Economy and Competitiveness (MINECO) under the grant AYA2011-22614. We thank D. A. Garc\'{\i}a-Hern\'andez, M. Lugaro, and O. Zamora for fruitful discussions and C. Morisset for including Ni in {\sc PyNeb}. We also thank the NEBULATOM school and its organizers (C. Mendoza, C. Morisset, and G. Stasi\'nska) because it is the place where this idea was born. Part of the computations presented in this paper were obtained using hardware from projects CONACyT-CB-153985 and UNAM-PAPIIT-107215. This paper is partially based on previously unpublished data obtained at the European Southern Observatory, Chile, with proposals ESO 68.C-0149(A) and 070.C-0008(A).


\bsp	
\label{lastpage}

\begin{thebibliography}{}

\bibitem[\protect\citeauthoryear{Barlow}{1978}]{bar78} 
Barlow M.~J., 1978, MNRAS, 183, 417

\bibitem[\protect\citeauthoryear{{Bautista}}{{Bautista}}{2001}]{bautista01}
{Bautista} M.~A.,  2001, \aap, 365, 268

\bibitem[\protect\citeauthoryear{Bautista \& Pradhan}{1996}]{bau96} 
Bautista M.~A., Pradhan A.~K., 1996, A\&AS, 115, 551 

\bibitem[\protect\citeauthoryear{Bensby, Feltzing, \& Oey}{2014}]{ben14} 
Bensby T., Feltzing S., Oey M.~S., 2014, A\&A, 562, A71

\bibitem[\protect\citeauthoryear{Busso, Gallino, \& Wasserburg}{1999}]{bus99} 
Busso M., Gallino R., Wasserburg G.~J., 1999, ARA\&A, 37, 239

\bibitem[\protect\citeauthoryear{{Butler} \& {Zeippen}}{{Butler} \&
  {Zeippen}}{1989}]{ButlerZeippen89}
{Butler} K.,  {Zeippen} C.~J.,  1989, \aap, 208, 337

\bibitem[\protect\citeauthoryear{Clyne et al.}{2015}]{cly15} 
Clyne N., Akras S., Steffen W., Redman M., Goncalves D.~R., Harvey E., 2015, A\&A, in press

\bibitem[\protect\citeauthoryear{{Delgado-Inglada}, {Morisset} \& {Stasi{\'n}ska}}{{Delgado-Inglada} et~al.}{2014}]{del14a}
{Delgado-Inglada} G.,  {Morisset} C.,    {Stasi{\'n}ska} G.,  2014, MNRAS, 440, 536

\bibitem[\protect\citeauthoryear{Delgado-Inglada \& Rodr{\'{\i}}guez}{2014}]{del14b} 
Delgado-Inglada G., Rodr{\'{\i}}guez M., 2014, ApJ, 784, 173 

\bibitem[\protect\citeauthoryear{{Delgado-Inglada}, {Rodr{\'\i}guez}, {Mampaso} \& {Viironen}}{{Delgado-Inglada} et~al.}{2009}]{delgadoingladaetal09}
{Delgado Inglada} G.,  {Rodr{\'\i}guez} M.,  {Mampaso} A.,    {Viironen} K.,
  2009, \apj, 694, 1335

\bibitem[\protect\citeauthoryear{Delgado-Inglada et al.}{2015}]{del15} 
Delgado-Inglada G., Rodr{\'{\i}}guez M., Peimbert M., Stasi{\'n}ska G., Morisset C., 2015, MNRAS, 449, 1797

\bibitem[\protect\citeauthoryear{Draine}{1990}]{dra90} 
Draine B.~T., 1990, ASPC, 12, 193 

\bibitem[\protect\citeauthoryear{Draine}{2009}]{dra09} 
Draine B.~T., 2009, ASPC, 414, 453 

\bibitem[\protect\citeauthoryear{Esteban, Garc{\'{\i}}a-Rojas, \& P{\'e}rez-Mesa}{2015}]{est15} 
Esteban C., Garc{\'{\i}}a-Rojas J., P{\'e}rez-Mesa V., 2015, MNRAS, 452, 1553

\bibitem[\protect\citeauthoryear{{Esteban}, {Peimbert}, {Garc{\'{\i}}a-Rojas},
  {Ruiz}, {Peimbert} \& {Rodr{\'{\i}}guez}}{{Esteban}
  et~al.}{2004}]{estebanetal04}
{Esteban} C.,  {Peimbert} M.,  {Garc{\'{\i}}a-Rojas} J.,  {Ruiz} M.~T.,
  {Peimbert} A.,    {Rodr{\'{\i}}guez} M.,  2004, \mnras, 355, 229

\bibitem[\protect\citeauthoryear{{Esteban}, {Peimbert}, {Torres-Peimbert}, \& {Escalante}}{{Esteban} et~al.}{1998}]{est98}
{Esteban} C., {Peimbert} M., {Torres-Peimbert} S., {Escalante} V., 1998, \mnras, 295, 401

\bibitem[\protect\citeauthoryear{{Fang} \& {Liu}}{{Fang} \&
  {Liu}}{2011}]{fangliu11}
{Fang} X.,  {Liu} X.-W.,  2011, \mnras, 415, 181

\bibitem[\protect\citeauthoryear{Field}{1974}]{fie74} 
Field G.~B., 1974, ApJ, 187, 453 

\bibitem[\protect\citeauthoryear{{Froese Fischer} \& {Tachiev}}{{Froese
  Fischer} \& {Tachiev}}{2004}]{froesefischertachiev04}
{Froese Fischer} C.,  {Tachiev} G.,  2004, Atomic Data and Nuclear Data Tables,
  87, 1

\bibitem[\protect\citeauthoryear{Galavis, Mendoza, \& Zeippen}{1997}]{galavis97} 
Galavis M.~E., Mendoza C., Zeippen C.~J., 1997, A\&AS, 123, 159

\bibitem[\protect\citeauthoryear{{Garc{\'\i}a-Rojas}, {Esteban}, {Peimbert}, {Rodr{\'{\i}}guez}, {Peimbert} \& {Ruiz}}{{Garc{\'\i}a-Rojas} et~al.}{2007}]{garciarojasetal07}
{Garc{\'\i}a-Rojas} J.,  {Esteban} C.,  {Peimbert} A.,  {Rodr{\'{\i}}guez} M.,
  {Peimbert} M.,    {Ruiz} M.~T.,  2007, Revista Mexicana de Astronomia y
  Astrofisica, 43, 3

\bibitem[\protect\citeauthoryear{{Garc{\'\i}a-Rojas}, {Esteban}, {Peimbert}, {Rodr{\'{\i}}guez}, {Ruiz} \& {Peimbert}}{{Garc{\'\i}a-Rojas}
  et~al.}{2004}]{garciarojasetal04}
{Garc{\'\i}a-Rojas} J.,  {Esteban} C.,  {Peimbert} M.,  {Rodr{\'{\i}}guez} M.,
  {Ruiz} M.~T.,    {Peimbert} A.,  2004, \apjs, 153, 501

\bibitem[\protect\citeauthoryear{Garc{\'{\i}}a-Rojas et al.}{2015}]{garciarojasetal15} 
Garc{\'{\i}}a-Rojas J., Madonna S., Luridiana V., Sterling N.~C., Morisset C., Delgado-Inglada G., Toribio San Cipriano L., 2015, MNRAS, 452, 2606 

\bibitem[\protect\citeauthoryear{Garc{\'{\i}}a-Rojas et al.}{2013}]{gar13} 
Garc{\'{\i}}a-Rojas J., Pe{\~n}a M., Morisset C., Delgado-Inglada G., Mesa-Delgado A., Ruiz M.~T., 2013, A\&A, 558, A122 

\bibitem[\protect\citeauthoryear{{Garc{\'\i}a-Rojas}, {Pe{\~n}a}, {Morisset}, {Mesa-Delgado} \& {Ruiz}}{{Garc{\'\i}a-Rojas}
  et~al.}{2012}]{garciarojasetal12}
{Garc{\'\i}a-Rojas} J.,  {Pe{\~n}a} M.,  {Morisset} C.,  {Mesa-Delgado} A.,
  {Ruiz} M.~T.,  2012, \aap, 538, A54

\bibitem[\protect\citeauthoryear{Henry}{1984}]{hen84} 
Henry R.~B.~C., 1984, ApJ, 281, 644

\bibitem[\protect\citeauthoryear{Herwig}{2005}]{her05} 
Herwig F., 2005, ARA\&A, 43, 435

\bibitem[\protect\citeauthoryear{Hirashita}{2012}]{hir12} 
Hirashita H., 2012, MNRAS, 422, 1263

\bibitem[\protect\citeauthoryear{{Jenkins}}{{Jenkins}}{2009}]{jen09}
{Jenkins} E.~B.  2009, \apj, 700, 1299

\bibitem[\protect\citeauthoryear{{Johansson}, {Zethson}, {Hartman}, {Ekberg},
  {Ishibashi}, {Davidson} \& {Gull}}{{Johansson}
  et~al.}{2000}]{Johanssonetal00}
{Johansson} S.,  {Zethson} T.,  {Hartman} H.,  {Ekberg} J.~O.,  {Ishibashi} K.,
   {Davidson} K.,    {Gull} T.,  2000, \aap, 361, 977

\bibitem[Jones \& Nuth(2011)]{jones11} 
Jones, A.~P., \& Nuth, J.~A.\ 2011, \aap, 530, A44 

\bibitem[\protect\citeauthoryear{{Jones} \& {Nuth}}{{Jones} \& {Nuth}}{2011}]{jonesnuth11}
{Jones} A.~P.,  {Nuth}  J.~A.,  2011, \aap, 530, A44

\bibitem[\protect\citeauthoryear{Karakas \& Lattanzio}{2014}]{kar14} 
Karakas A.~I., Lattanzio J.~C., 2014, PASA, 31, e030

\bibitem[\protect\citeauthoryear{{Kisielius}, {Storey}, {Ferland} \&
  {Keenan}}{{Kisielius} et~al.}{2009}]{kisieliusetal09}
{Kisielius} R.,  {Storey} P.~J.,  {Ferland} G.~J.,    {Keenan} F.~P.,  2009,
  \mnras, 397, 903

\bibitem[\protect\citeauthoryear{Krueger \& Czyzak}{1970}]{kru70} 
Krueger T.~K., Czyzak S.~J., 1970, RSPSA, 318, 531

\bibitem[\protect\citeauthoryear{Liu et al.}{2000}]{liu00} 
Liu X.-W., Storey P.~J., Barlow M.~J., Danziger I.~J., Cohen M., Bryce M., 
2000, MNRAS, 312, 585 

\bibitem[\protect\citeauthoryear{{Lodders}}{{Lodders}}{2010}]{lod10}
{Lodders} K.,  2010, in Principles and Perspectives in Cosmochemistry, 
ed. A. Goswami \& B.~E. Reddy (Berlin: Springer), 379

\bibitem[\protect\citeauthoryear{{Lucy}}{{Lucy}}{1995}]{lucy95}
{Lucy} L.~B.,  1995, A\&A, 294, 555

\bibitem[\protect\citeauthoryear{{Luridiana}, {Morisset} \& {Shaw}}{{Luridiana}
  et~al.}{2015}]{luridianaetal15}
{Luridiana} V.,  {Morisset} C.,    {Shaw} R.~A.,  2015, \aap, 573, A42

\bibitem[\protect\citeauthoryear{{Mendoza}}{{Mendoza}}{1982}]{men82}
{Mendoza} C.,  1982, J. Phys. B, 15, 867

\bibitem[\protect\citeauthoryear{{Mendoza} \& {Zeippen}}{{Mendoza} \&
  {Zeippen}}{1982}]{mendozazeippen82a}
{Mendoza} C.,  {Zeippen} C.~J.,  1982, \mnras, 198, 127

\bibitem[\protect\citeauthoryear{{Mesa-Delgado}, {Esteban},
  {Garc{\'{\i}}a-Rojas}, {Luridiana}, {Bautista}, {Rodr{\'{\i}}guez},
  {L{\'o}pez-Mart{\'{\i}}n} \& {Peimbert}}{{Mesa-Delgado}
  et~al.}{2009}]{mesadelgadoetal09b}
{Mesa-Delgado} A.,  {Esteban} C.,  {Garc{\'{\i}}a-Rojas} J.,  {Luridiana} V.,
  {Bautista} M.,  {Rodr{\'{\i}}guez} M.,  {L{\'o}pez-Mart{\'{\i}}n} L.,
  {Peimbert} M.,  2009, \mnras, 395, 855

\bibitem[\protect\citeauthoryear{Morard et al.}{2013}]{mor13} 
Morard G., Siebert J., Andrault D., Guignot N., Garbarino G., Guyot F., Antonangeli D., 2013, E\&PSL, 373, 169

\bibitem[\protect\citeauthoryear{Nomoto, Kobayashi, \& Tominaga}{2013}]{nom13} 
Nomoto K., Kobayashi C., Tominaga N., 2013, ARA\&A, 51, 457

\bibitem[\protect\citeauthoryear{Osterbrock, Tran, \& Veilleux}{1992}]{ost92} 
Osterbrock D.~E., Tran H.~D., Veilleux S., 1992, ApJ, 389, 305 

\bibitem[\protect\citeauthoryear{{Podobedova}, {Kelleher} \&
  {Wiese}}{{Podobedova} et~al.}{2009}]{Podobedovaetal09}
{Podobedova} L.~I.,  {Kelleher} D.~E.,    {Wiese} W.~L.,  2009, Journal of
  Physical and Chemical Reference Data, 38, 171

\bibitem[\protect\citeauthoryear{{Porter}, {Ferland}, {Storey} \&
  {Detisch}}{{Porter} et~al.}{2012}]{porteretal12}
{Porter} R.~L.,  {Ferland} G.~J.,  {Storey} P.~J.,    {Detisch} M.~J.,  2012, MNRAS, 425, L28

\bibitem[\protect\citeauthoryear{{Porter}, {Ferland}, {Storey} \&
  {Detisch}}{{Porter} et~al.}{2013}]{porteretal13}
{Porter} R.~L.,  {Ferland} G.~J.,  {Storey} P.~J.,    {Detisch} M.~J.,  2013, MNRAS, 433, L89

\bibitem[\protect\citeauthoryear{{Quinet}}{{Quinet}}{1996}]{quinet96}
{Quinet} P.,  1996, \aaps, 116, 573

\bibitem[\protect\citeauthoryear{{Ramsbottom}, {Bell} \& {Keenan}}{{Ramsbottom}
  et~al.}{1997}]{Ramsbottometal97}
{Ramsbottom} C.~A.,  {Bell} K.~L.,    {Keenan} F.~P.,  1997, \mnras, 284, 754

\bibitem[\protect\citeauthoryear{Rechy-Garc{\'{\i}}a, Pe{\~n}a, \& Garc{\'{\i}}a-Rojas}{2014}]{rec14} 
Rechy-Garc{\'{\i}}a J., Pe{\~n}a M., Garc{\'{\i}}a-Rojas J., 2014, apn6.conf, 78 

\bibitem[\protect\citeauthoryear{Reddy, Lambert, \& Allende Prieto}{2006}]{red06} 
Reddy B.~E., Lambert D.~L., Allende Prieto C., 2006, MNRAS, 367, 1329

\bibitem[\protect\citeauthoryear{Reddy et al.}{2003}]{red03} 
Reddy B.~E., Tomkin J., Lambert D.~L., Allende Prieto C., 2003, MNRAS, 340, 304 

\bibitem[\protect\citeauthoryear{Rodr{\'{\i}}guez}{1999}]{rod99} 
Rodr{\'{\i}}guez M., 1999, A\&A, 348, 222 

\bibitem[\protect\citeauthoryear{{Rodr{\'{\i}}guez} \& {Rubin}}{{Rodr{\'{\i}}guez} \& {Rubin}}{2005}]{rodriguezrubin05}
{Rodr{\'{\i}}guez} M.,  {Rubin} R.~H.,  2005, ApJ, 626, 900

\bibitem[\protect\citeauthoryear{{Sharpee}, {Williams}, {Baldwin} \& {van
  Hoof}}{{Sharpee} et~al.}{2003}]{sharpeeetal03}
{Sharpee} B.,  {Williams} R.,  {Baldwin} J.~A.,    {van Hoof} P.~A.~M.,  2003,
  \apjs, 149, 157

\bibitem[\protect\citeauthoryear{Sharpee et al.}{2007}]{sha07} 
Sharpee B., Zhang Y., Williams R., Pellegrini E., Cavagnolo K., Baldwin J.~A., Phillips M., Liu X.-W., 2007, ApJ, 659, 1265

\bibitem[\protect\citeauthoryear{Sterling \& Dinerstein}{2008}]{ste08} 
Sterling N.~C., Dinerstein H.~L., 2008, ApJS, 174, 158 

\bibitem[\protect\citeauthoryear{{Sterling}, {Porter} \&
  {Dinerstein}}{{Sterling} et~al.}{2015}]{sterlingetal15}
{Sterling} N.~C.,  {Porter} R.~L.,  {Dinerstein} H.~L.,  2015, ApJ, 218, 25

\bibitem[\protect\citeauthoryear{{Storey} \& {Hummer}}{{Storey} \& {Hummer}}{1995}]{storeyhummer95}
{Storey} P.~J.,  {Hummer} D.~G.,  1995, MNRAS, 272, 41

\bibitem[\protect\citeauthoryear{{Storey}, {Sochi} \& {Badnell}}{{Storey} et~al.}{2014}]{storeyetal14}
{Storey} P.~J.,  {Sochi} T.,    {Badnell} N.~R.,  2014, \mnras, 441, 3028

\bibitem[\protect\citeauthoryear{{Tayal}}{{Tayal}}{2011}]{tayal11}
{Tayal} S.~S.,  2011, \apjs, 195, 12

\bibitem[\protect\citeauthoryear{{Tayal} \& {Zatsarinny}}{{Tayal} \& {Zatsarinny}}{2010}]{TayalZatsarinny10}
{Tayal} S.~S.,  {Zatsarinny} O.,  2010, \apjs, 188, 32

\bibitem[\protect\citeauthoryear{{Tsamis}, {Barlow}, {Liu}, {Danziger} \&
  {Storey}}{{Tsamis} et~al.}{2003}]{tsamisetal03b}
{Tsamis} Y.~G.,  {Barlow} M.~J.,  {Liu} X.-W.,  {Danziger} I.~J.,    {Storey}
  P.~J.,  2003, \mnras, 345, 186

\bibitem[\protect\citeauthoryear{Verner et al.}{2000}]{ver00} 
Verner E.~M., Verner D.~A., Baldwin J.~A., Ferland G.~J., Martin P.~G., 
2000, ApJ, 543, 831 

\bibitem[\protect\citeauthoryear{Whittet}{2003}]{whi03} 
Whittet, D. C. B. 2003, Dust in the Galactic Environment, 2nd ed. (Bristol: IOP)

\bibitem[\protect\citeauthoryear{{Zhang}}{{Zhang}}{1996}]{zhang96}
{Zhang} H.,  1996, \aaps, 119, 523

\bibitem[\protect\citeauthoryear{{Zhang} \& {Liu}}{{Zhang} \& {Liu}}{2002}]{zhangliu02}
{Zhang} Y.,  {Liu} X.-W.,  2002, \mnras, 337, 499

\bibitem[\protect\citeauthoryear{{Zhang} \& {Liu}}{{Zhang} \& {Liu}}{2006}]{zhangliu06}
{Zhang} Y.,  {Liu} X.-W.,  2006, in IAU Symp. 234. Eds. M.~J. Barlow \& R.~H. M\'endez, p. 547

\bibitem[\protect\citeauthoryear{{Zhang}, {Liu}, {Liu} \& {Rubin}}{{Zhang} et~al.}{2005}]{zhangetal05}{Zhang} Y.,  {Liu} X.-W.,  {Liu} Y.,    {Rubin} R.~H.,  2005, \mnras, 358, 457

\end{thebibliography}
\end{document}